\documentclass[a4paper]{amsart}
\usepackage{fullpage}

    \usepackage{amsfonts,amssymb,amsmath}
    \usepackage{amsmath,amsthm}
    \usepackage[mathscr]{euscript}
    \usepackage{dsfont,bbm}
    \allowdisplaybreaks 

            \usepackage[usenames,dvipsnames]{color}
            \usepackage{soul}
            \sethlcolor{cyan}

            \newtheorem*{myTheorem}{Factorization Theorem}
            \newcommand{\myProof}{\noindent \emph{Proof}.\ }
            \newtheorem{myLemma}{Lemma}

    \usepackage[dvips]{pstricks}
    \psset{xunit=.45pt,yunit=.45pt,runit=.45pt}

\def\sumstar{\mathop{\sum\nolimits^{*}}}
\def\sumstars{\mathop{\sum\nolimits^{**}}}  

\begin{document}

\title{On factorizing $F$-matrices in $\mathcal{Y}(sl_n)$ and $\mathcal{U}_q(\widehat{sl_n})$ spin chains}

\author{S G M\textsuperscript{c}Ateer and M Wheeler}

\address{Department of Mathematics and Statistics,
                 The University of Melbourne,
                 Parkville, Victoria, Australia.}
\email{s.mcateer@ms.unimelb.edu.au, m.wheeler@ms.unimelb.edu.au}

\begin{abstract}
We consider quantum spin chains arising from $N$-fold tensor products of the fundamental evaluation representations of $\mathcal{Y}(sl_n)$ and $\mathcal{U}_q(\widehat{sl_n})$. Using the partial $F$-matrix formalism from the seminal work of Maillet and Sanchez de Santos, we derive a completely factorized expression for the $F$-matrix of such models and prove its equivalence to the expression obtained by Albert, Boos, Flume and Ruhlig. A new relation between the $F$-matrices and the Bethe eigenvectors of these spin chains is given. 
\end{abstract}

\maketitle

\setcounter{section}{0}

\section{Introduction}
\label{sec:introduction}

The first application of Drinfel'd twists \cite{dri1,dri2,dri3} in the context of the algebraic Bethe Ansatz was by Maillet and Sanchez de Santos in \cite{ms}.   In this seminal paper, the authors considered a representation of the Drinfel'd twist -- which they called the {\it $F$-matrix} -- on a tensor product of finite-dimensional irreducible modules of the underlying quantum affine algebra. Letting $V_i$ denote such an irreducible module for all $1\leq i \leq N$, the $F$-matrix $F_{1\dots N} \in {\rm End}(V_1 \otimes \cdots \otimes V_N)$ was defined as the solution\footnote{All solutions of (\ref{fact-intro}) are related by elementary transformations, so we say that this equation admits a unique solution.} of the equation
\begin{align}
F_{\sigma(1) \dots \sigma(N)}
R^{\sigma}_{1\dots N}
=
F_{1\dots N}
\label{fact-intro}
\end{align}
for all permutations $\sigma \in S_N$, where $R^{\sigma}_{1\dots N}$ is a specific product of the $R$-matrices associated with the quantum affine algebra in question. An important result of \cite{ms} was an explicit construction of $F_{1\dots N}$ in the particular case of the algebras $\mathcal{Y}({sl_2})$ and $\mathcal{U}_q(\widehat{sl_2})$, which later found use in the algebraic Bethe Ansatz approach to the XXX and XXZ spin-$\frac{1}{2}$ chains, respectively \cite{kmt}.

An extension of the results in \cite{ms} to algebras of higher rank was obtained shortly afterwards by Albert, Boos, Flume and Ruhlig in \cite{abfr}.   In this work, the authors obtained a summation formula for $F_{1\dots N}$ satisfying (\ref{fact-intro}), in the case where the $R$-matrix is of the type corresponding to $\mathcal{Y}({sl_n})$, and went on to study the Bethe eigenvectors of the higher rank XXX spin chains under the change of basis induced by the $F$-matrix. What is missing in \cite{abfr} is a construction of $F_{1\dots N}$ using the partial $F$-matrix approach developed in \cite{ms}. Moreover it is natural to expect that the $sl(n)$ factorizing problem can be solved by a reduction, in $(n-2)$ steps, to the $sl(2)$ factorizing problem whose solution is known \cite{ms}. Such a method is in keeping with the spirit of the nested Bethe Ansatz \cite{kr,br}, which is the technique used to construct the eigenvectors of these models. Indeed, in the nested Bethe Ansatz approach to the $sl(n)$ spin chains, the eigenvectors of the transfer matrix are obtained via $(n-2)$ reductions to the $sl(2)$ problem, the solution of which is known from the algebraic Bethe Ansatz \cite{kbi}. 

The purpose of this paper is to settle the questions raised in the previous paragraph. Our main result is a new formula for the $F$-matrix $F_{1\dots N}$ for $N$-fold tensor products of the fundamental evaluation representations of $\mathcal{Y}({sl_n})$ and $\mathcal{U}_q(\widehat{sl_n})$.  We will show that the $F$-matrix admits the completely factorized expression
\begin{align}
F_{1\dots N}
=
\left\{
\begin{array}{ll}
F^{2}_{1 \dots N}
\dots
F^{n-1}_{N \dots 1}
F^{n}_{1\dots N},
&
\quad n \text{ even},
\\ \\
F^{2}_{N \dots 1}
\dots
F^{n-1}_{N \dots 1}
F^{n}_{1\dots N},
&
\quad n \text{ odd},
\end{array}
\right.
\label{F-intro}
\end{align}
where each $F^k_{1\dots N}$ has an analogous form to the $F$-matrices of \cite{ms}, and is composed of a product of \emph{partial $F$-matrices} $F^k_{1 \dots (i-1),i}$ as follows
\begin{equation}
    F^k_{1 \dots  N} 
    = 
    F^k_{1,2} F^k_{12,3} \dots F^k_{1 \dots (N-1), N}.
\end{equation}
 A key feature of our work will be the notion of \emph{tiers.} Throughout the paper we say that each $F^k_{1\dots N}$ is situated at tier-$k$ in reference to the fact that it depends on the interaction of only $k$ state variables, or in other words, has $sl(k)$ type behaviour. To prove that the $F$-matrix (\ref{F-intro}) satisfies the factorizing equation (\ref{fact-intro}) we proceed on a tier-by-tier basis, establishing an $sl(k)$ version of this identity for all $2\leq k \leq n$. This achieves the aim mentioned in the last paragraph. As we will show, the expression (\ref{F-intro}) turns out to be equivalent to the result obtained in \cite{abfr}, despite the fact that our construction is quite different.

In our recent paper \cite{mw} we presented a review of \cite{ms}, working in terms of a new diagrammatic notation motivated by the six-vertex model. The cornerstone of our approach was a diagrammatic representation of the partial $F$-matrices used in \cite{ms}. In this paper we generalize our previous notation \cite{mw}, to allow a diagrammatic description of the $sl(n)$ $F$-matrix (\ref{F-intro}). For clarity of exposition, we will present algebraic and diagrammatic versions of almost every equation.

In Section 2 we collect a number of definitions which are used throughout this work. These include the $sl(n)$ type $R$-matrix, its reduction to $sl(k)$ (which we call the tier-$k$ $R$-matrix), and the tier-$k$ partial and complete $F$-matrices. In keeping with our previous paper \cite{mw}, we define these objects both algebraically and in terms of diagrammatic tensor notation similar to that of Penrose  \cite{pen}. In Section 3 we give our expression for the $F$-matrix (\ref{F-intro}), and prove that it satisfies the factorizing equation (\ref{fact-intro}). The proof is very transparent, since it relies only on two simple identities involving the tier-$k$ partial $F$-matrices. Section 4 contains examples of our formula (\ref{F-intro}) in the special cases of $sl(2)$ and $sl(3)$. The $sl(2)$ case is included for completeness, so that the reader can compare with the original work in \cite{ms}. The $sl(3)$ case illustrates how the components of the tensor $F_{1\dots N}$ can be extracted in general, and allows us to explain the equivalence of our formula to that obtained in \cite{abfr}. 

In Section 5 we study other properties of the $F$-matrix which were proved in \cite{abfr}, namely its lower triangularity and invertibility. Section 6 contains a review of the nested Bethe Ansatz expression for the eigenvectors of the $\mathcal{Y}(sl_n)$ and $\mathcal{U}_q(\widehat{sl_n})$ spin chains. We derive new formulae relating these eigenvectors with the $F$-matrices studied earlier in the paper.

\section{Definitions and expression for $F$-matrix}
\label{sec:definitions}

\subsection{Preliminary remarks on notation}
\label{ssec:remarks}
In all instances we use $n$ in reference to the Lie algebra $sl(n)$, while $N$ is used for the length of the spin chain in the models being studied. 

We will consider many different tensors of varying rank. The building block of all these tensors is the $n\times n$ \emph{elementary matrix} $E^{(kl)}$ which acts in the vector space $V = \mathbb{C}^n$, and whose components are given by
\begin{align}
    ( E^{(kl)} )_i^j = \delta_{ik} \delta_{jl}.
    \label{elem}
\end{align}
More complicated tensors $T_{1\dots N}$ are formed by taking linear combinations of tensor products of the elementary matrices (\ref{elem}). Here we use the subscript $1\dots N$ to indicate that $T_{1\dots N}$ acts in $V_1 \otimes \cdots \otimes V_N$, where each $V_1,\dots,V_N$ is a copy of $\mathbb{C}^n$, and it informs us that $T_{1\dots N}$ is a rank $2N$ tensor of complex dimension $n$. In general we will use the notation 
\begin{align}
    (T_{1\dots N})_{i_1\dots i_N}^{j_1\dots j_N}
\end{align}
to indicate the components of $T_{1\dots N}$, where each index takes values in the set $\{1,\dots,n\}$.    

It will be our practice to omit certain dependences, where they are unnecessary in the context.  For example, in the case of the $sl(n)$ $R$-matrix (defined in Subsection \ref{subsec:R}) we will write $R_{12}$ instead of $R_{12}(u_1, u_2)$, since the dependence on $u_1,u_2$ is already implied by the subscript. A similar convention will apply to tensors of higher rank.

\subsection{$R$-matrix, Yang-Baxter and unitarity equation}
\label{subsec:R}
In this paper we consider quantum spin chains based on fundamental evaluation representations of $\mathcal{Y}(sl_n)$ and $\mathcal{U}_q(\widehat{sl_n})$. The \emph{$R$-matrix} for these models \cite{jim2} is of the form
\begin{multline}
    R_{12}(u_1,u_2)
    =
    a(u_1-u_2)
    \sum_{1 \leq i \leq n} 
    E^{(ii)}_1 E^{(ii)}_2
    +
    b(u_1-u_2)
    \sum_{\substack{1 \leq i,j \leq n \\ i \not= j}}
    E^{(ii)}_1 E^{(jj)}_2
    \\
    +
    \sum_{1 \leq i<j \leq n}
    \left(
    c_{+}(u_1-u_2)
    E^{(ij)}_1 E^{(ji)}_2
    +
    c_{-}(u_1-u_2)
    E^{(ji)}_1 E^{(ij)}_2
    \right)
    \label{Rmat}
\end{multline}
where $E^{(ij)}_1$, $E^{(ij)}_2$ denote the elementary matrices (\ref{elem}) acting in the respective vector spaces $V_1$, $V_2$, and $u_1, u_2$ are the \emph{rapidities} associated with those spaces.  The \emph{weight functions} $a, b, c_{\pm}$ are given by
\begin{align}
    a(u) = 1, 
    \quad
    b(u) = \frac{u}{u + \eta},
    \quad
    c_{\pm}(u) = \frac{\eta}{u + \eta} 
\end{align}
in the case of $\mathcal{Y}(sl_n)$, and by
\begin{align}
    a(u) = 1,
    \quad
    b(u) 
    = \frac{\sinh u}{\sinh(u + \eta)},
    \quad
    c_{\pm}(u) 
    = \frac{e^{\pm u}\sinh\eta}{\sinh(u + \eta)}
\end{align}
in the case of $\mathcal{U}_q(\widehat{sl_n})$, where $\eta$ is the \emph{crossing parameter} of the model. In order to encompass both of these cases, we will simply refer to (\ref{Rmat}) as the \emph{$sl(n)$ $R$-matrix,} and make no further reference to the particular values of the weight functions. 

Often we will find it more natural to talk about the components of the $sl(n)$ $R$-matrix, rather than its tensorial form. From the definition (\ref{Rmat}) we find that the components are given by
\begin{align}
    (R_{12})^{j_1 j_2}_{i_1 i_2} =
    \left\{
    \begin{array}{ll}
        a(u_1 - u_2),   
        & \quad i_1 = i_2 = j_1 = j_2, \\ \\
        b(u_1 - u_2),   
        & \quad i_1 = j_1,\ i_2 = j_2,\ i_1 \neq i_2, \\ \\
        c_{+}(u_1 - u_2), 
        & \quad i_1 = j_2,\ i_2 = j_1,\ i_1 < i_2, \\ \\
        c_{-}(u_1 - u_2), 
        & \quad i_1 = j_2,\ i_2 = j_1,\ i_1 > i_2, \\ \\
        0,
        & \quad \text{otherwise}.
    \end{array}
    \right.
    \label{Rcomp}
\end{align}
Notice that for clarity we omit the dependence on the rapidity variables when discussing the components of tensors. We will represent the components of the $R$-matrix diagrammatically as a pair of  intersecting lines with indices, as follows
\begin{align}
    (R_{12})^{j_1 j_2}_{i_1 i_2}\ =\ \input{diagrams/define-r}\ .
    \label{r-vert}
\end{align}
In this diagram, the top half comprises the \emph{arms} of the vertex and the bottom half comprises the \emph{legs} of the vertex. Collectively we call them the \emph{limbs} of the vertex. The indices $\{i_1,i_2,j_1,j_2\}$ will be called  \emph{colours}. In general, each limb of a vertex is assigned one of the colours $\{1,\dots,n\}$. As is apparent from (\ref{Rcomp}), the two colours on the arms must be the same as the two colours on the legs, or else the vertex has weight zero.

Whenever we draw an $R$-vertex within a larger diagram, we need a systematic way of identifying which limbs comprise its arms and which comprise its legs. We do this by ensuring that any given line in the diagram (associated with a space $V_k$) will have one end terminating at the bottom of the diagram (corresponding to the index $i_k$) and one end terminating at the top of the diagram (corresponding to the index $j_k$). This induces a bottom-to-top orientation on every line, and hence a bottom-to-top orientation on every vertex, fixing its arms and legs in the same way as described above. To translate between the algebraic and diagrammatic versions of a tensor, we assume that left-to-right multiplication corresponds with bottom-to-top contraction in a diagram. 

At times when we draw $R$-vertices, we may omit indices from the limbs. The meaning of this omission depends on context. If the limb is connected to the limb of another vertex, the omission of the index implies that it is to be summed over all values $\{1,\dots,n\}$. If the limb is external to the diagram, the omission of the index simply means that we have no interest in its value.  For example in the case of
\begin{align}
     (R_{13} R_{23})_{i_1 i_2 i_3}^{j_1 j_2 j_3} 
     =
     \sum_{k_3=1}^n(R_{13})_{i_1 i_3}^{j_1 k_3} (R_{23})_{i_2 k_3}^{j_2 j_3} 
     =
     \ \input{diagrams/example-contraction}\ ,
\end{align}
the index on the line connecting the vertices is omitted because it is constrained to the summation and plays no role.
    
\begin{myLemma}\label{lem:yb}
    (Yang-Baxter and unitarity equation.)
    \begin{align}
        R_{23}R_{13}R_{12} &= R_{12}R_{13}R_{23},
        \label{YB}
        \\
        R_{21} R_{12} &= I_{12}.
        \label{unit} 
    \end{align}
    in diagrammatic notation, the components of these relations may be written as
    \begin{align}
        \input{diagrams/diag-yb-1}\ &=\ \input{diagrams/diag-yb-2}\ ,
        \label{YB-diag}
        \\
        \input{diagrams/diag-unitarity-1}\ &=\ \input{diagrams/diag-unitarity-2}\ .
        \label{unit-diag} 
    \end{align}
\end{myLemma}

\myProof Both of these relations can be proved by comparing the components of the tensors on the left and right hand sides, and establishing each as a scalar identity. For more information on these equations and their role in exactly solvable models we refer the reader to \cite{bax1,jim3,wda}. \hfill$\square$

\subsection{Identity matrix }

The \emph{identity matrix} $I_{12}$ is given by
\begin{align}
    I_{12}
    =
    \sum_{1 \leq i,j \leq n}
    E^{(ii)}_1 E^{(jj)}_2.
\end{align}
Alternatively, in component form, we have
\begin{align}
    (I_{12})^{j_1 j_2}_{i_1 i_2} 
    = 
    \delta_{i_1 j_1} \delta_{i_2 j_2}.
\end{align}
We will represent the components of the identity matrix diagrammatically as a pair of lines which \emph{do not} intersect,
\begin{align}
    (I_{12})^{j_1 j_2}_{i_1 i_2}\ =\ \input{diagrams/define-i}\ .
\end{align}

\subsection{$R^{\sigma}_{1\dots N}$ as a bipartite graph}

Let $\sigma\{1,\ldots,N\} = \{\sigma(1),\ldots,\sigma(N)\}$ be an arbitrary permutation of the set of integers $\{1,\ldots,N\}$. A standard device is to represent this permutation as a bipartite graph. This is achieved by writing down two rows of integers $\{N,\ldots,1\}$ and $\{\sigma(N),\ldots,\sigma(1)\}$, the former directly above the latter, and connecting each integer $i$ in the top row with $i$ in the bottom row.  The only constraints on the graph are that no three lines may intersect at a point and that no line may self-intersect.  We denote the resulting graph by $G(\sigma)$. 

Using the diagrammatic representation (\ref{r-vert}) of the $R$-matrix, we define $R^{\sigma}_{1\ldots N}$ to be the rank $2N$ tensor corresponding to the graph $G(\sigma)$. That is, its components $(R^{\sigma}_{1\dots N})_{i_1 \dots i_N}^{j_1 \dots j_N}$ are given by affixing the rows of indices $\{i_{\sigma(N)},\dots,i_{\sigma(1)}\}$ to the bottom and $\{j_N,\dots,j_1\}$ to the top of the graph $G(\sigma)$.  For example, when $N=5$ and $\sigma = \{5,2,4,1,3\}$ we have
\begin{align}
    (R^{\sigma}_{1\dots 5})_{i_1 \dots i_5}^{j_1 \dots j_5} =\ \input{diagrams/define-r-sigma-example}
\end{align}
which may be expanded in tensor notation as
\begin{align}
    R^{\sigma}_{1\dots 5} = R_{25} R_{45} R_{31} R_{35} R_{15} R_{13} R_{14} R_{34} R_{12}.
\end{align}
Generally there are many ways of drawing $G(\sigma)$, giving rise to different intersections between the lines. However, all ways of drawing $G(\sigma)$ are equivalent up to applications of the unitarity (\ref{unit-diag}) and Yang-Baxter equation (\ref{YB-diag}), so there is no ambiguity in the definition.

The definition of the tensor $R^{\sigma}_{1\dots N}$ is  crucial to the remainder of the paper. The problem considered by Maillet and Sanchez de Santos in \cite{ms} is to construct an invertible matrix $F_{1\dots N}$ satisfying 
\begin{align}
    R^{\sigma}_{1\dots N}
    =
    F_{\sigma(1) \dots \sigma(N)}^{-1}
    F_{1\dots N}
    \label{factorize}
\end{align}
for all permutations $\sigma \in S_N$, giving rise to the terminology {\it factorizing $F$-matrix.} To do this we will seek a solution $F_{1\dots N}$ of the equation (\ref{fact-intro}) before showing (in Subsection \ref{sec:inverse}) that $F_{1\dots N}$ is invertible. This is in agreement with the approach used in \cite{abfr} for the $\mathcal{Y}(sl_n)$ spin chains. The new feature of our work is an independent expression for $F_{1\dots N}$, which will be shown equivalent to that obtained in \cite{abfr} in Subsection \ref{gen-to-sln}, using the following lemma. 

\begin{myLemma}\label{lem:bip}
    Let $\{i_1,\dots,i_N\}$ be any set of integers taking values in $\{1,\dots,n\}$ and let
    $\sigma,\rho$ be any two permutations of $\{1,\dots,N\}$ which satisfy
    \begin{align}
    i_{\sigma(1)} \leq \cdots \leq i_{\sigma(N)},
    \quad
    i_{\rho(1)} \leq \cdots \leq i_{\rho(N)}.
    \label{order}
    \end{align}
    Then we have the following equivalence between the components of the graphs 
    $R^{\sigma}_{1\dots N}$ and $R^{\rho}_{1\dots N}$,
    \begin{align}
    (R^{\sigma}_{1\dots N})^{j_1 \dots j_N}_{i_1 \dots i_N}
    =
    (R^{\rho}_{1\dots N})^{j_1\dots j_N}_{i_1 \dots i_N}.
    \end{align}
\end{myLemma}

\myProof Up to applications of the unitarity and Yang-Baxter equation, the graphs corresponding to $\sigma$ and $\rho$ only differ from one another in the ordering of the spaces at the base of the diagram. Furthermore, thanks to the assumption (\ref{order}), they only differ within groups of consecutive  identical colours. Now we observe that $E^{(ii)}_1 E^{(ii)}_2 R_{12} = E^{(ii)}_1 E^{(ii)}_2 I_{12}$, for all $1\leq i \leq n$ (this arises due to colour conservation and the chosen normalization of the $R$-matrix). The lemma is proved by repeatedly applying this relation to consecutive pairs of spaces with identical colour in the graph for $\sigma$, until the graph for $\rho$ is produced. \hfill $\square$

This method of proof may be illustrated by consideration of an example. For $N=5$ and $i_2 = i_4 = i_5 = 1$, $i_1 = i_3 = 2$, two permutations which achieve the required ordering (\ref{order}) are $\sigma = \{5,2,4,1,3\}$ and $\rho = \{2,4,5,1,3\}$. We then find that 
\begin{equation}
(R^{\sigma}_{1\dots 5})_{i_1 \dots i_5}^{j_1 \dots j_5}
=
    \input{diagrams/example-reorder-1}\
    =\ \input{diagrams/example-reorder-2}\ 
    =\ \input{diagrams/example-reorder-3}\ 
    =
(R^{\rho}_{1\dots 5})_{i_1 \dots i_5}^{j_1 \dots j_5}\ ,
\end{equation}
where for transparency we have indicated colours in brackets below the relevant indices. The first equality between diagrams uses the relation $E^{(11)}_2 E^{(11)}_5 R_{25} = E^{(11)}_2 E^{(11)}_5 I_{25}$, then $E^{(11)}_4 E^{(11)}_5 R_{45} = E^{(11)}_4 E^{(11)}_5 I_{45}$. The second equality between diagrams is simply a rearrangement of the non-intersecting lines.

\subsection{Tier-$k$ version of $R$-matrix and Yang-Baxter equation}
Another definition which is essential to the remainder of the paper is that of the \emph{tier-$k$ $R$-matrix}. This effectively constitutes a reduction of the $sl(n)$ $R$-matrix to $sl(k)$. Such a reduction is characteristic of the nested Bethe Ansatz approach to the $sl(n)$ XXX and XXZ spin chains and is therefore natural in the construction of the $F$-matrices for the same models. For all $1 \leq k \leq n$ we define the tier-$k$ $R$-matrix $R^k_{12}$ as
\begin{align}
    R_{12}^{k}
    =
    I_{12}
    +
    \sum_{1 \leq i,j \leq k}
    E^{(ii)}_1 E^{(jj)}_2
    (R_{12}-I_{12}).
    \label{def:tier-k}
\end{align}
The effect of this definition is that $R^k_{12}$ behaves like an identity matrix in the presence of any colours greater than $k$, and like an ordinary $sl(n)$ $R$-matrix otherwise. The special cases $k=1$ and $k=n$ warrant further mention. Since $E^{(11)}_1 E^{(11)}_2 (R_{12}-I_{12}) = 0$ we clearly have $R^1_{12} = I_{12}$. Furthermore, since $\sum_{1\leq i,j \leq n} E^{(ii)}_1 E^{(jj)}_2 = I_{12}$ we see that $R^n_{12} = R_{12}$.

In component form, we have an even simpler understanding of the tier-$k$ $R$-matrix,
\begin{align}
    ( R_{12}^k )_{i_1 i_2}^{j_1 j_2}
    =
    \left\{ 
    \begin{array}{ll}
    (R_{12})_{i_1 i_2}^{j_1 j_2},
    & \quad
    i_1 \leq k\ {\rm and}\ i_2 \leq k,
    \\
    \\
    (I_{12})_{i_1 i_2}^{j_1 j_2},
    & \quad
    i_1 > k\ {\rm or}\ i_2 > k.
    \end{array}
    \right.
    \label{tier-k-comp}
\end{align}
We will represent the components of the tier-$k$ $R$-matrix diagrammatically in the same way as the $R$-matrix itself, except it will bear an additional label, as follows
\begin{align}
    (R^k_{12})^{j_1 j_2}_{i_1 i_2}\ =\          \input{diagrams/define-tier-k-r}\ .
    \label{dia:def-tier-k}
\end{align}
Using equation (\ref{tier-k-comp}) we may also realize the right hand side of (\ref{dia:def-tier-k}) in terms of existing diagrams. In particular, we observe that 
\begin{align}
    \input{diagrams/define-tier-k-r}\ =
    \left\{
        \begin{array}{ll}
            \input{diagrams/define-r}\ , & \quad i_1 \leq k \text{ and } i_2 \leq k, \\ \\ 
            \input{diagrams/define-i}\ , & \quad i_1 > k \text{ or } i_2 > k. \\
        \end{array}
    \right.
\end{align}

\begin{myLemma}\label{lem:tier-k-yb}
    (Tier-$k$ version of the Yang-Baxter and unitarity equation.) 
    \begin{align}
    R^k_{23}R^k_{13}R^k_{12} &= R^k_{12}R^k_{13}R^k_{23}, 
    \label{tier-k-YB} \\
    R^k_{21} R^k_{12} &= I_{12}.
    \label{tier-k-unitarity}
    \end{align}
\end{myLemma}

\myProof We write the components of the Yang-Baxter equation (\ref{tier-k-YB}) and the unitarity relation (\ref{tier-k-unitarity}) diagrammatically as
\begin{align}
    \input{diagrams/diag-tier-k-yb-1}\ &=\ \input{diagrams/diag-tier-k-yb-2}\ , \\
    \input{diagrams/diag-tier-k-unitarity-2}\ &=\ \input{diagrams/diag-tier-k-unitarity-1}\ ,
\end{align}
respectively.

To establish the tier-$k$ Yang-Baxter equation, we must consider two cases.  \underline{\emph{Case 1.}} All the incoming colours $\{i_1,i_2,i_3\}$ are less than or equal to $k$.  In this case the tier-$k$ $R$-matrices behave as ordinary $R$-matrices, so the relation holds by virtue of Lemma \ref{lem:yb}.  \underline{\emph{Case 2.}} At least one incoming colour, is greater than $k$. In this case the vertices carrying that colour become identities and the relation becomes a trivial equality of tier-$k$ $R$-matrices. For example, when $i_3>k$ we have
\begin{align}
    \input{diagrams/proof-tier-k-yb-1}\ &=\ \input{diagrams/proof-tier-k-yb-2}\ .
\end{align}
The tier-$k$ unitarity equation is established by considering similar cases. \hfill$\square$

\subsection{Tier-$k$ partial $F$-matrix}
Let the \emph{tier-$k$ monodromy matrix} $R^k_{1 \dots (N-1),N}$ be the tensor formed by contraction of  $(N-1)$ tier-$k$ $R$-matrices on the vector space $V_N$, 
\begin{align}
    R^k_{1 \dots (N-1), N} =
    R^k_{1N} R^k_{2N} \dots R^k_{(N-1)N}.
    \label{monod}
\end{align}
In accordance with our comments on subscripts in Subsection \ref{ssec:remarks}, 
$R^k_{1 \dots (N-1),N}$ acts in $V_1 \otimes \cdots \otimes V_N$. Diagrammatically we may represent a contraction by simply joining an arm of one tier-$k$ $R$-matrix to a leg of another. Thereby, we may write the components of tier-$k$ monodromy matrix diagrammatically as 
\begin{align}
    (R^k_{1 \dots (N-1), N})^{j_1 \dots j_N}_{i_1 \dots i_N} 
    &\ \input{diagrams/define-tier-k-mono-1}\ .
\end{align}
Here we have not written indices on the internal lines of this diagram, since on these lines a summation is implied over all colours $\{1,\dots,n\}$. Indeed in this case and in many of the cases that follow, it would be cumbersome and unnecessary to assign indices to such internal lines.

Similarly to above, let a \emph{string of identity matrices} $I_{1 \dots (N-1), N}$ be the tensor formed by contraction of $(N-1)$ identity matrices on the vector space $V_N$,
\begin{align}
    I_{1 \dots (N-1), N} =
    I_{1N} I_{2N} \dots I_{(N-1)N}.
    \label{string}
\end{align}
We may write the components of this string diagrammatically as 
\begin{align}
    (I^k_{1 \dots (N-1), N})^{j_1 \dots j_N}_{i_1 \dots i_N} &\ \input{diagrams/define-i-mono-1}\ ,
\end{align}
in which the lines do not intersect.  In this tensor there is no interaction between the spaces.

Having fixed the definitions (\ref{monod}) and (\ref{string}), we now introduce the \emph{tier-$k$ partial $F$-matrix} $F^k_{1 \dots (N-1),N}$ which is central to our construction of the factorizing $F$-matrix. We define it as
\begin{align}
    F^k_{1 \dots (N-1), N} =
    E^{(kk)}_N I_{1 \dots (N-1), N} 
    + 
    \sum_{\substack{1 \leq i \leq n \\ i \not= k}} E^{(ii)}_N R^k_{1 \dots (N-1), N}.
\end{align}
This definition extends that of the $sl(2)$ partial $F$-matrix, given in \cite{ms}, to algebras of higher rank. Indeed, by taking $n=k=2$ we recover the definition found in \cite{ms}. The components of $F^k_{1\dots (N-1),N}$ fall into two categories, depending on the value of the index $i_N$,
\begin{align}
    (F^k_{1\dots (N-1), N})^{j_1\dots j_N}_{i_1 \dots i_N}
    =
    \left\{
    \begin{array}{ll}
    ( I_{1\dots (N-1),N} )^{j_1 \dots j_N}_{i_1 \dots i_N},
    & \quad
    i_N = k,
    \\
    \\
    ( R^k_{1\dots (N-1),N} )^{j_1 \dots j_N}_{i_1 \dots i_N},
    & \quad
    i_N \not= k.
    \end{array}
    \right.
    \label{partial-comp}
\end{align}
In diagrammatic notation we will represent the components of the tier-$k$ partial $F$-matrix as 
\begin{align}
    (F^k_{1 \dots (N-1), N})^{j_1 \dots j_N}_{i_1 \dots i_N} 
    &\ =\ \input{diagrams/define-tier-k-partial-f-1}\ ,
\end{align}
where in view of (\ref{partial-comp}) we may write
\begin{align}
    \input{diagrams/diag-define-tier-k-partial-f-1}\ =
    \left\{
        \begin{array}{ll}
            \input{diagrams/diag-define-tier-k-partial-f-2}\ , 
            & \quad  i_N = k, \\ \\ 
            \input{diagrams/diag-define-tier-k-partial-f-3}\ , 
            & \quad  i_N \neq k. \\
        \end{array}
    \right.
    \label{eqn:tier-k-partial-f}
\end{align}
On the diagram for the tier-$k$ partial $F$-matrix, the box containing the symbol $k$ represents sensitivity to the colour entering on that line. If the colour is equal to $k$, then it is a string of identity matrices, otherwise it is a tier-$k$ monodromy matrix. In proving identities involving the tier-$k$ partial $F$-matrices, we will often find it useful break the proof into two cases, namely that where the colour at the box is equal to $k$ and that where the colour at the box is not equal to $k$.

Note that in equation \eqref{eqn:tier-k-partial-f} we have omitted all of the indices which were bystanders in the definition. On occasion we will omit such indices for the sake of clarity.

\subsection{Tier-$k$ $F$-matrix} 
Proceeding in analogy with \cite{ms}, the \emph{tier-$k$ $F$-matrix} $F^k_{1\dots N}$ is constructed as a product of $(N-1)$ tier-$k$ partial $F$-matrices.  We define
\begin{align}
    F^k_{1 \dots N} =
    F^k_{1, 2} F^k_{1 2, 3} \dots F^k_{1 \dots (N-2), (N-1)} F^k_{1 \dots (N-1), N}.
    \label{def:tier-k-f}
\end{align}
We may write the components of the tier-$k$ $F$-matrix in diagrammatic notation as 
\begin{align}
    (F^k_{1 \dots  N})^{j_1 \dots j_N}_{i_1 \dots i_N}\ =\ \input{diagrams/define-tier-k-f-1}\ .
\end{align}
 Once again we remark that the internal lines in this diagram are assumed to be summed over all colours $\{1,\dots,n\}$. Hence the diagrammatic way of writing the components of $F^k_{1\dots N}$ is much more convenient than using purely symbolic notation, which would require introducing $N(N-2)$ dummy indices to encode the summations.

Notice that $F^k_{1\dots N}$ does not act symmetrically on the quantum spaces 
$V_1 \otimes \cdots \otimes V_N$. Therefore it is sometimes of interest to consider the same object, but with a permuted action on the quantum spaces. To this end, for all permutations $\sigma$ of $\{1,\dots,N\}$ the tensor $F^k_{\sigma(1) \dots \sigma(N)}$ is assumed to be same operator but now acting on $V_{\sigma(1)} \otimes \cdots \otimes V_{\sigma(N)}$. In Subsection \ref{ssec:Fmat}, we will be particularly interested by the permutation $\sigma\{1,\dots, N\} = \{N,\dots,1\}$ which reverses the order of the quantum spaces.

\subsection{Factorizing $F$-matrix}
\label{ssec:Fmat}
Using the definitions made up to this point, we now give an expression for the solution of the factorizing equation (\ref{factorize}), the \emph{$F$-matrix.} It is constructed as a product of tier-$k$ $F$-matrices, where $k$ ranges over $\{2,\dots,n\}$.  We define
\begin{align}
    F_{1 \dots N} = 
    \left\{ \begin{array}{ll}
        F^2_{1 \dots N} F^3_{N \dots 1} \dots F^{n-1}_{N \dots 1} F^n_{1 \dots N}, 
        & \quad n \text{ even}, 
        \\ \\
        F^2_{N \dots 1} F^3_{1 \dots N} \dots F^{n-1}_{N \dots 1} F^n_{1 \dots N}, 
        & \quad n \text{ odd}. \\
    \end{array} \right.
    \label{def:f-mat}
\end{align}
We remark that the ordering of the quantum spaces is reversed at each step from tier-$k$ to tier-$(k-1)$. This explains why the definition of the $F$-matrix depends on the parity of $n$. In the special case $n=2$, this formula collapses to the expression of \cite{ms} for the $sl(2)$ models. 

Combining our previous diagrams, we can write the components of the $F$-matrix in diagrammatic notation as a chain of tier-$k$ $F$-matrices.  We have 
\begin{align}
    (F_{1 \dots  N})^{j_1 \dots j_N}_{i_1 \dots i_N} = \input{diagrams/define-f-2}\ .
\end{align}
Here the dotted lines are used to demarcate the tiers, and the order of indices at the base of the diagram depends on the parity of $n$. In the case $n=2$ we have only a single tier, and this diagram specializes to the one obtained in \cite{mw}.

From its definition, it is straightforward to see that tier-$k$ of the $F$-matrix only admits non-trivial interaction between the colours $\{1,\dots, k\}$. Any line bearing a colour greater than $k$ will simply peel away from this part of the lattice. Hence we say that tier-$k$ has a reduced, $sl(k)$ type behaviour. This decomposition of the $F$-matrix into structures which are reduced iteratively, is reminiscent of the nested Bethe Ansatz approach to the $sl(n)$ spin chains \cite{kr,br}. We review the nested Bethe Ansatz in Section 6 and make this correspondence more concrete, by showing that the $F$-matrices and the Bethe eigenvectors of these models are explicitly linked.

As a final comment, in the case $n=2$ the $F$-matrix obeys the recursion relation
\begin{align}
    F_{1\dots N}
    =
    F_{1\dots (N-1)}
    F^{2}_{1 \dots (N-1), N}
\end{align}
in which all action in the quantum space $V_N$ comes via the partial $F$-matrix $F^{2}_{1 \dots (N-1), N}$. This recursion allows an inductive proof of formulae for the {\it twisted} monodromy matrix operators\footnote{The twist of the operator $O_{1\dots N}$ is defined to be $F_{1\dots N} O_{1\dots N} F^{-1}_{1\dots N}$.} in the $\mathcal{Y}(sl_2)$ and $\mathcal{U}_q(\widehat{sl_2})$ models as in \cite{ms}. Unfortunately, as is easily checked, a similar recursion relation does not exist in the cases $n \geq 3$. This makes it harder to prove formulae for the twisted operators of the higher rank models \cite{abfr}, but we shall not be concerned with this problem in this paper. 

\begin{myTheorem}
The $F$-matrix (\ref{def:f-mat}) satisfies the factorizing equation
    \begin{align}
    F_{\sigma(1)\dots \sigma(N)}
    R^{\sigma}_{1\dots N}
    =
    F_{1\dots N}
    \label{factorize2}
    \end{align}
with respect to bipartite graph $R^{\sigma}_{1\dots N}$, for all permutations $\sigma$. The next section is devoted to the proof of this result. 
\end{myTheorem}

\section{Proof of Factorization Theorem}
\label{sec:proof}

The proof proceeds by establishing lemmas for passing individual $R$-matrices firstly through tier-$k$ partial $F$-matrices, then tier-$k$ $F$-matrices, and finally the $F$-matrix itself.

\subsection{Two lemmas involving tier-$k$ partial $F$-matrices}

We establish two lemmas for passing tier-$k$ $R$-matrices through tier-$k$ partial $F$-matrices.  The first lemma involves a tier-$k$ $R$-matrix which is \emph{not} at the leftmost position,  namely, \emph{not} $R^k_{(N-1)N}$.

\begin{myLemma}\label{lem:a} For all $2 \leq k \leq n$ and $ 1 \leq i < N-1$ we have
\begin{align} 
    F^k_{1\dots (i+1) i \dots (N-1), N} R^k_{i(i+1)} 
    = 
    R^k_{i(i+1)} F^k_{1\dots (N-1),N}.
    \label{result1}
\end{align} 
\end{myLemma}

\myProof
We present an entirely diagrammatic proof. In diagrammatic form, the proposed equation (\ref{result1}) becomes
\begin{align}
    \input{diagrams/diag-lemma-1-1}\ =\  \input{diagrams/diag-lemma-1-2}\ .
\end{align}
Notice that the tier-$k$ $R$-vertex on the left hand side \emph{cannot} be attached to the leftmost two lines, ensuring that the condition $i<N-1$ is satisfied. Let us now consider the possible values of the index $i_N$ occurring at the position of the box. There are two cases which require separate treatment, as we describe below. In either case, the first and last equality is due to the definition (\ref{eqn:tier-k-partial-f}) of the tier-$k$ partial $F$-matrix.

\underline{\emph{Case 1, $i_N=k$.}} 
We use the top part of the definition (\ref{eqn:tier-k-partial-f}) to obtain
\begin{align}
    \input{diagrams/lemma-1-k-case-1}\ &=\  \input{diagrams/lemma-1-k-case-2} \\ \nonumber
    &=\ \input{diagrams/lemma-1-k-case-3} \\ \nonumber
    &=\ \input{diagrams/lemma-1-k-case-4}\ .
\end{align}
The first and last equalities follow from the definition of the tier-$k$ partial $F$-matrix (\ref{partial-comp}) and the second equality is trivial.

\underline{\emph{Case 2, $i_N\neq k$.}}
We use the bottom part of the definition (\ref{eqn:tier-k-partial-f}) to obtain
\begin{align}
    \input{diagrams/lemma-1-not-k-case-1}\ &=\ \input{diagrams/lemma-1-not-k-case-2} \\ \nonumber
    &=\ \input{diagrams/lemma-1-not-k-case-3} \\ \nonumber
    &=\ \input{diagrams/lemma-1-not-k-case-4}\ .
\end{align}
In this case we have used the tier-$k$ Yang-Baxter equation (\ref{tier-k-YB}) to achieve the second equality.\hfill$\square$

The second lemma involves a tier-$k$ $R$-matrix which \emph{is} at the leftmost position, namely, $R^k_{(N-1)N}$.

\begin{myLemma}\label{lem:b} For all $2 \leq k \leq n$ we have
\begin{align}
    F^k_{1 \dots (N-2), N} 
    F^k_{1 \dots (N-2) N, (N-1)} 
    R^k_{(N-1)N} 
    = 
    R^{k-1}_{N(N-1)} 
    F^k_{1 \dots (N-2), (N-1)} 
    F^k_{1 \dots (N-1), N}.
    \label{result2}
\end{align}
\end{myLemma}

\myProof Again we give a diagrammatic proof.  In diagrammatic form the proposed equation (\ref{result2}) becomes\footnote{Here and in several subsequent diagrams, lines will be displayed using red and blue purely for added clarity where colour viewing is available.} 
\begin{align}
    \input{diagrams/diag-lemma-2-1-col}\ =\ \input{diagrams/diag-lemma-2-2-col}\ .
\end{align}
We emphasize again that the tier-$k$ $R$-vertex on the left hand side \emph{is} attached to the leftmost two lines. From here we divide the proof into four cases, corresponding to the possible values of the indices $i_{N-1},i_N$ situated at the position of the boxes. In each case, the first and last equality is due to the definition (\ref{eqn:tier-k-partial-f}) of the tier-$k$ partial $F$-matrix.

\underline{\emph{Case 1, $i_{N-1}=k,\ i_N=k$.}}
Applying the top part of the definition (\ref{eqn:tier-k-partial-f}) to both the $i_{N-1}$ and $i_N$ lines, we obtain
\begin{align}
    \input{diagrams/lemma-2-kk-case-1}\ &=\ \input{diagrams/lemma-2-kk-case-2} \\ \nonumber
    &=\ \input{diagrams/lemma-2-kk-case-3} \\ \nonumber
    &=\ \input{diagrams/lemma-2-kk-case-4} \\ \nonumber
    &=\ \input{diagrams/lemma-2-kk-case-5}\ .
\end{align}
The second equality follows from the fact $E^{(kk)}_1 E^{(kk)}_2 R^k_{12} = E^{(kk)}_1 E^{(kk)}_2 I_{12}$, while the third equality is a trivial rearrangement of identity matrices.

\underline{\emph{Case 2, $i_{N-1}\neq k,\ i_N\neq k$.}}
Applying the bottom part of the definition (\ref{eqn:tier-k-partial-f}) to both the $i_{N-1}$ and $i_N$ lines, we obtain
\begin{align}
    \input{diagrams/lemma-2-nknk-case-1}\ &=\ \input{diagrams/lemma-2-nknk-case-2} \\ \nonumber
    &=\ \input{diagrams/lemma-2-nknk-case-3} \\ \nonumber
    &=\ \input{diagrams/lemma-2-nknk-case-4}\ .
\end{align}
The second equality is achieved by repeated application of the tier-$k$ Yang-Baxter equation (Lemma~\ref{lem:tier-k-yb}).

\underline{\emph{Case 3, $i_{N-1} \neq k,\ i_N = k$.}}
Applying the bottom part of the definition (\ref{eqn:tier-k-partial-f}) to the $i_{N-1}$ line and the top part of the definition (\ref{eqn:tier-k-partial-f}) to the $i_N$ line, we obtain
\begin{align}
    \input{diagrams/lemma-2-knk-case-1}\ &=\ \input{diagrams/lemma-2-knk-case-2} \\ \nonumber
    &=\ \input{diagrams/lemma-2-knk-case-3} \\ \nonumber
    &=\ \input{diagrams/lemma-2-knk-case-4} \\ \nonumber
    &=\ \input{diagrams/lemma-2-knk-case-5}\ .
\end{align}
The second equality is achieved by application of the tier-$k$ unitarity relation, while the third equality is a trivial rearrangement of the position of the identity matrices.

\underline{\emph{Case 4, $i_{N-1} = k,\ i_N \neq k$.}}
Applying the top part of the definition (\ref{eqn:tier-k-partial-f}) to the $i_{N-1}$ line and the bottom part of the definition (\ref{eqn:tier-k-partial-f}) to the $i_N$ line, we obtain
\begin{align}
    \input{diagrams/lemma-2-nkk-case-1}\ &=\ \input{diagrams/lemma-2-nkk-case-2} \\ \nonumber
    &=\ \input{diagrams/lemma-2-nkk-case-3} \\ \nonumber
    &=\ \input{diagrams/lemma-2-nkk-case-4} \\ \nonumber
    &=\ \input{diagrams/lemma-2-nkk-case-5}\ .
\end{align}
The second equality follows from a trivial rearrangement of identity matrices, while the third equality is simply a redrawing of the diagram.\hfill$\square$

\subsection{A lemma involving the tier-$k$ $F$-matrix}

The following lemma, a corollary of the previous two, allows us to pass a tier-$k$ $R$-matrix through a tier-$k$ $F$-matrix. Loosely speaking, this lemma allows us to deconstruct the $sl(n)$ factorizing problem into a series of reductions from $sl(k)$ to $sl(k-1)$, for all values of $k$. 

\begin{myLemma}\label{lem:c}
For all $1 \leq i < N$ and $2 \leq k \leq n$ we have
\begin{align}
    F^k_{1 \dots (i+1) i \dots N} R^k_{i(i+1)} = R^{k-1}_{(i+1)i} F^k_{1 \dots N}.
\end{align}
\end{myLemma}

\myProof The tier-$k$ $R$-matrix entering at the top of the diagram may be translated vertically through the lattice, by $N-i-1$ applications of Lemma~\ref{lem:a}. Then a single application of Lemma~\ref{lem:b} causes the extraction of a tier-$(k-1)$ $R$-matrix from the base of the lattice.
\hfill$\square$

For example, by setting $N=7$ and $i=3$, we obtain
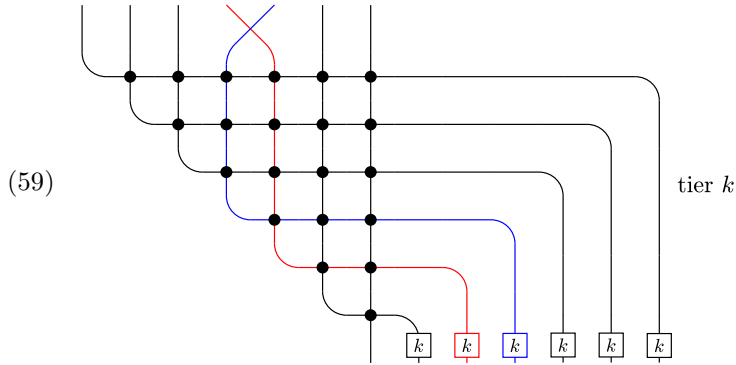
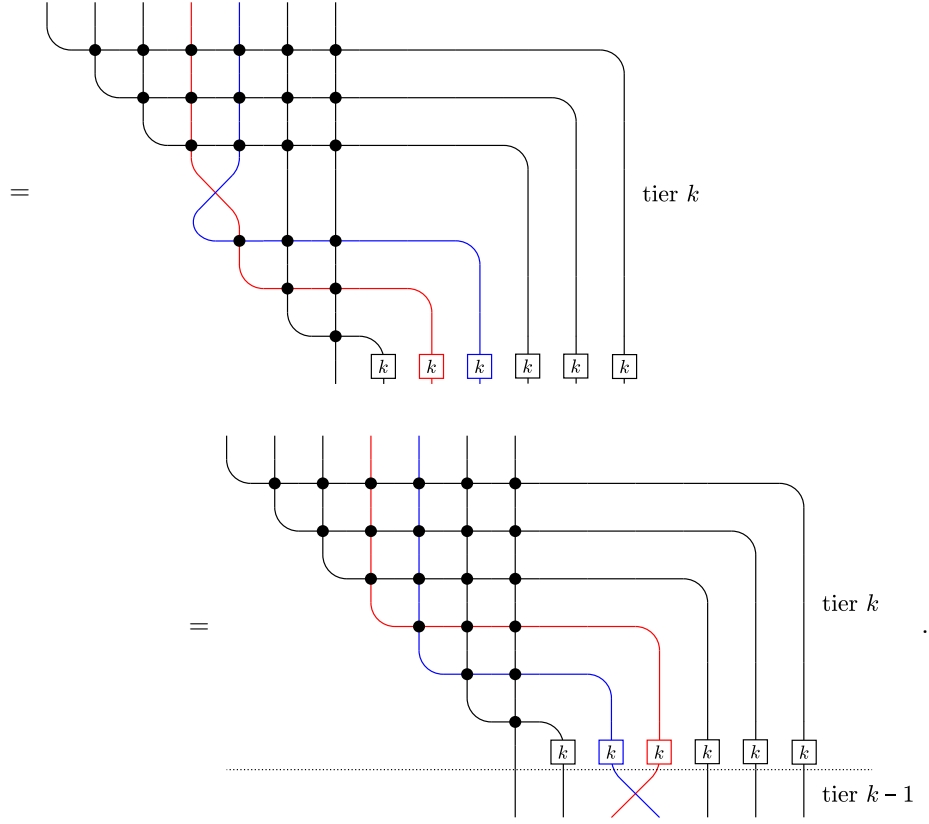
\begin{multline}
    \input{diagrams/example-lemma-3-1-col} \\ \\
    =\ \input{diagrams/example-lemma-3-2-col} \\ \\
    =\ \input{diagrams/example-lemma-3-3-col}\ . 
\end{multline}
The first equality is achieved by $N-i-1=3$ applications of Lemma~\ref{lem:a}, and the second equality is achieved by one application of Lemma~\ref{lem:b}. Notice that as a result of this procedure the order of the two participating lattice lines is reversed.

\subsection{Proof of Factorization Theorem for individual $R$-matrices}

This lemma, a corollary of the previous one, allows us to pass an individual $R$-matrix through the $F$-matrix. When viewed as permutations, the individual $R$-matrices correspond to adjacent site-swaps which form a generating set for the set all permutations.  

\begin{myLemma}\label{lem:d}
For all $1 \leq i < N$ we have
\begin{align}
F_{1 \dots (i+1) i \dots N} R_{i(i+1)} = F_{1 \dots N}.
\end{align}
\end{myLemma}

\myProof We firstly recall that $R_{12} = R_{12}^{n}$, from the definition of the tier-$n$ $R$-matrix (\ref{def:tier-k}). Then by using the expression (\ref{def:f-mat}) for the $F$-matrix we have
\begin{align}
F_{1 \dots (i+1) i \dots N}
R_{i(i+1)}
=
\left\{
\begin{array}{ll}
F_{1 \dots (i+1) i \dots N}^{2}
\dots 
F_{N \dots i (i+1) \dots 1}^{n-1}
F_{1 \dots (i+1) i \dots N}^{n}
R_{i(i+1)}^{n}, 
& 
\quad n \text{ even},
\\ \\
F_{N \dots i (i+1) \dots 1}^{2}
\dots 
F_{N \dots i (i+1) \dots 1}^{n-1}
F_{1 \dots (i+1) i \dots N}^{n}
R_{i(i+1)}^{n},
& 
\quad n \text{ odd}.
\end{array}
\right.
\end{align}
Now by making $(n-1)$ applications of Lemma~\ref{lem:c} we obtain
\begin{align}
F_{1 \dots (i+1) i \dots N}
R_{i(i+1)}
=
\left\{
\begin{array}{ll}
R_{(i+1)i}^{1}
F_{1 \dots N}^{2}
\dots 
F_{N \dots 1}^{n-1}
F_{1 \dots  N}^{n}, 
& 
\quad n \text{ even},
\\ \\
R_{i(i+1)}^{1}
F_{N \dots 1}^{2}
\dots 
F_{N \dots 1}^{n-1}
F_{1 \dots N}^{n},
& 
\quad n \text{ odd}.
\end{array}
\right.
\label{eq1}
\end{align}
Finally, using the fact $R^1_{12} = I_{12}$ we see that the tier-1 $R$-matrices on the right hand side of (\ref{eq1}) act as the identity. Then the right hand side matches the definition of the $F$-matrix (\ref{def:f-mat}) and the proof is complete.
\hfill$\square$ 

For example, by setting $n=3$, $N=7$ and $i=3$, we obtain
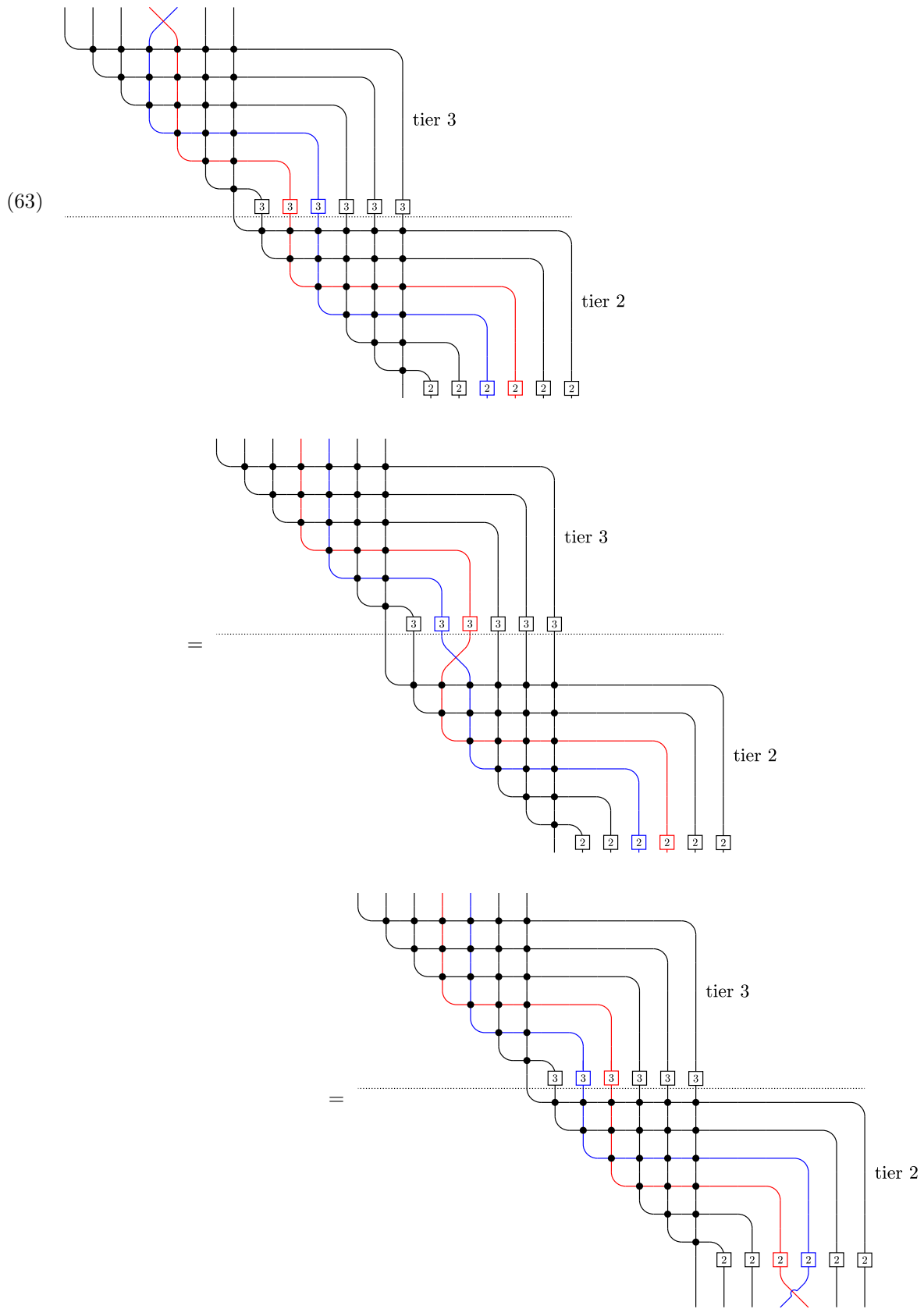
\begin{multline}
    \input{diagrams/example-lemma-4-1-col} \\ \\
    =\ \input{diagrams/example-lemma-4-2-col}  \\ \\
    =\ \input{diagrams/example-lemma-4-3-col}\ . 
\end{multline}
In the above, each equality results from a single application of Lemma \ref{lem:c}. In the final diagram we draw the tier-1 $R$-matrix at the base as an identity matrix.

\subsection{Proof of Factorization Theorem}

In the previous subsection we proved the Factorization Theorem for individual $R$-matrices. For all permutations $\sigma$ the bipartite graph $R^{\sigma}_{1\dots N}$ is just a composition of such $R$-matrices, and we obtain the general proof by repeated application of Lemma~\ref{lem:d}.
\hfill$\square$ 

\section{Examples}
\label{sec:examples}

In this section we present some examples which clarify the structure of the $sl(n)$ $F$-matrix (\ref{def:f-mat}). We will firstly consider the $sl(2)$ case of (\ref{def:f-mat}), when it reduces to the formula of \cite{ms}. Then we study in more detail the next simplest example, namely the $sl(3)$ specialization of (\ref{def:f-mat}). Finally for completeness we give the expression for the $sl(n)$ $F$-matrix as it appeared in \cite{abfr}, and show that our result is equivalent.

\subsection{$sl(2)$ $F$-matrix}

Taking the $n=2$ specialization of (\ref{def:f-mat}), we see that $F_{1\dots N}$ is equal to a single tier-2 $F$-matrix (\ref{def:tier-k-f}). Expanding this in terms of its tier-2 partial $F$-matrices, we recover the formula 
\begin{align}
F_{1\dots N}
=
F^{2}_{1,2}
F^{2}_{12,3}
\dots
F^{2}_{1\dots (N-1),N}
\end{align}
where we have defined
\begin{align}
F^{2}_{1\dots (i-1),i}
=
E^{(22)}_i 
I_{1\dots (i-1),i}
+
E^{(11)}_i
R^{2}_{1\dots (i-1),i}
\end{align}
for all $2 \leq i \leq N$. Recalling that in the $n=2$ case the tier-2 $R$-matrices satisfy $R^{2}_{12} = R_{12}$, this matches the expression for the $sl(2)$ $F$-matrix given in \cite{ms}. Diagrammatically, for $n=2$ we have
\begin{align}
    (F_{1\dots N})^{j_1\dots j_N}_{i_1\dots i_N} =\ \input{diagrams/example-sl2-f}
\end{align}
where the labeling of the tier is now redundant. Notice that this object behaves in the same way as the diagram defined in \cite{mw}, if each colour variable 1 is interchanged with an arrow pointing \emph{up} the page and each colour variable 2  is interchanged with an arrow pointing \emph{down} the page.

The labels $\{i_1,\dots,i_N\}$ at the base and $\{j_N,\dots,j_1\}$ at the top of the diagram indicate that we are dealing with the component $(F_{1\dots N})^{j_1\dots j_N}_{i_1\dots i_N}$ of the $F$-matrix. For a particular choice of $\{i_1,\dots,i_N\}$ it is useful to have a rule for expressing this component solely in terms of products of $R$-matrices. Starting from the label $i_2$ and progressing towards $i_N$, we use the following rule.

\emph{The colour $i_k$ can have the value 1 or 2. If $i_k=1$, the row of dotted vertices associated with $i_k$ becomes a row of $R$-matrices, and the base of this line remains stationary. If $i_k=2$, the row of dotted vertices associated with $i_k$ becomes a row of identity matrices. In this case these vertices become uncoupled, and the base of the line should be repositioned to the left of all other lines. Repeat for all $2 \leq k \leq N$.}

We illustrate this procedure with a simple example. For $n=2$, $N=7$ and $\{i_1,i_2,i_3,i_4,i_5,i_6,i_7\} = \{1,2,2,1,2,1,1\}$ we obtain
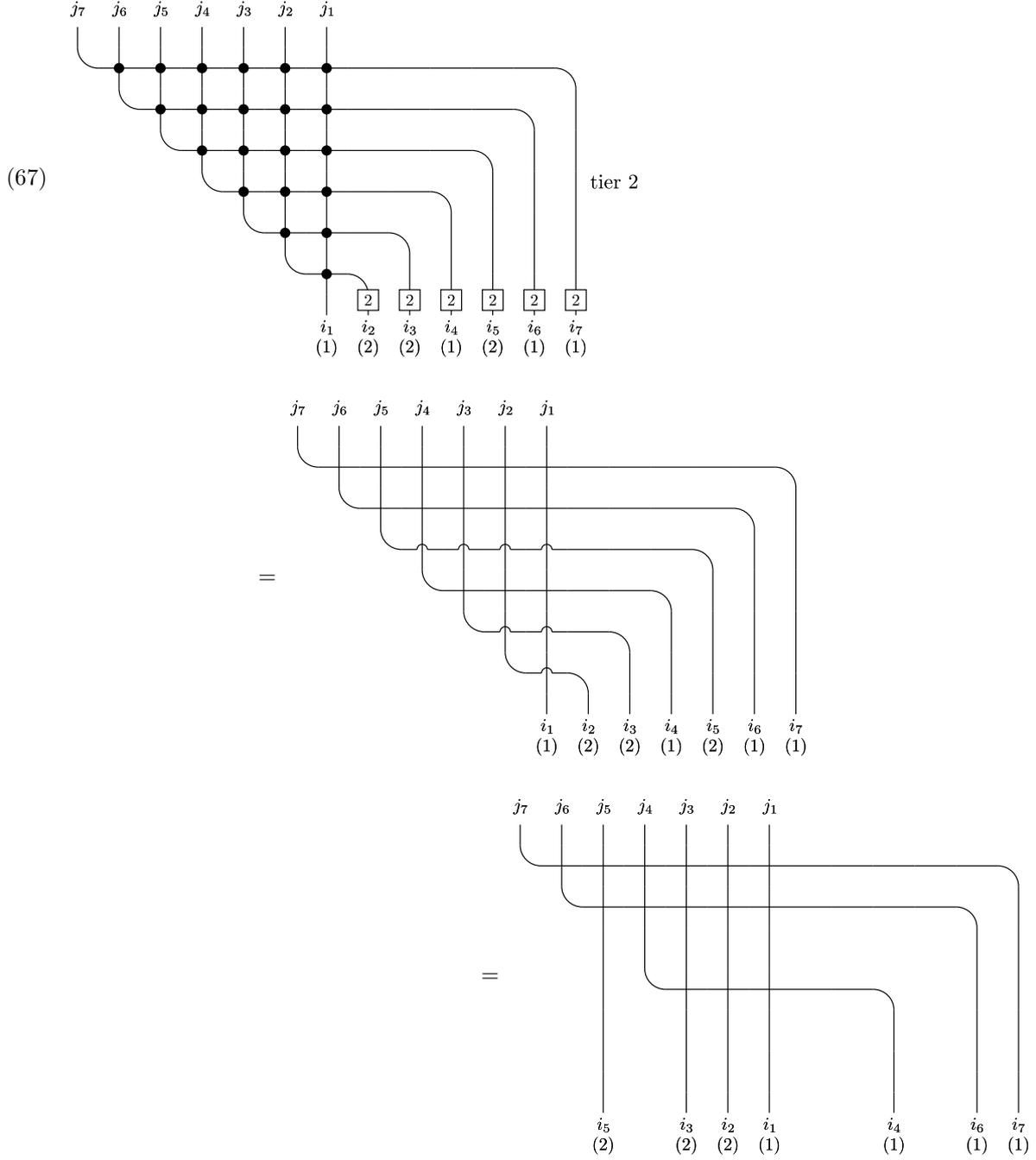
\begin{multline}
    \input{diagrams/example-sl2-sorting-1} \\ \\
    =\ \input{diagrams/example-sl2-sorting-2} \\ \\
    =\ \input{diagrams/example-sl2-sorting-3} 
\end{multline}
where the first equality is due to the definition of the tier-2 partial $F$-matrices \eqref{eqn:tier-k-partial-f}, and the second equality is obtained by trivial rearrangement of lines which do not intersect.

Notice that this simple algorithm ends by producing a bipartite graph, with incoming colours which are monotonically increasing from right to left. To this effect, let $\sigma$ be any permutation of $\{1,\dots,N\}$ for which 
$i_{\sigma(1)} \leq \cdots \leq i_{\sigma(N)}$. Considering the ordering properties of the algorithm described, it is apparent that
\begin{align}
(F_{1\dots N})^{j_1\dots j_N}_{i_1\dots i_N}
=
(R^{\sigma}_{1\dots N})^{j_1\dots j_N}_{i_1\dots i_N}.
\label{order2}
\end{align}
Generally, there are many permutations $\sigma$ which order the incoming colours as required. However, the formula (\ref{order2}) is independent of this freedom of choice for $\sigma$ by virtue of Lemma~\ref{lem:bip}.

\subsection{$sl(3)$ $F$-matrix}

For $n=3$, $F_{1\dots N}$ is a product of a tier-2 and tier-3 $F$-matrix (\ref{def:tier-k-f}). Expanding these in terms of their tier-2 and tier-3 partial $F$-matrices, we obtain
\begin{align}
F_{1\dots N}
=
F^{2}_{N,(N-1)}
F^{2}_{N (N-1),(N-2)}
\dots
F^{2}_{N \dots 2,1}
F^{3}_{1,2}
F^{3}_{12,3}
\dots
F^{3}_{1\dots (N-1),N}
\end{align}
where we have defined
\begin{align}
F^{2}_{N \dots (i+1),i}
&=
E^{(22)}_{i}
I_{N \dots (i+1),i}
+
\left(
E^{(11)}_i + E^{(33)}_i
\right)
R^{2}_{N \dots (i+1),i} ,
\\
F^{3}_{1 \dots (i-1),i}
&=
E^{(33)}_i
I_{1 \dots (i-1),i}
+
\left(
E^{(11)}_i + E^{(22)}_i
\right)
R^{3}_{1 \dots (i-1),i} ,
\end{align}
for all values of $i$. Diagrammatically, for $n=3$ we have
\begin{align}
    (F_{1\dots N})^{j_1\dots j_N}_{i_1\dots i_N} =\ \input{diagrams/example-sl3-f}\ .
\end{align}
As in the previous case, we introduce a rule for evaluating the components of the $F$-matrix in terms of products of $R$-matrices. This time we start at the label $i_{N}$ and progress towards $i_1$, as described below.

\emph{Each colour $i_k$ has an associated row of dotted vertices in tier-2 and in tier-3. The combined length of these rows is $N-1$. The colour $i_k$ can have value 1, 2 or 3. If $i_k = 1$, both the rows associated to this colour become rows of $R$-matrices, and the base of this line remains stationary. If $i_k = 2$, the associated row in tier-3 becomes a row of $R$-matrices, while the associated row in tier-2 becomes a row of identity matrices. The latter row decouples, and the base of the line should be repositioned to the left of all other lines bearing the colours 1 or 2. If $i_k = 3$, the line decouples completely from tier-2 and the associated row in tier-3 becomes a row of identity matrices. These also decouple, and the base of the line should be repositioned to the left of all other lines bearing the colours 1 or 2. Repeat for all $N \geq k \geq 1$.}

Again we give an example. For $n=3$, $N=7$ and $\{i_1,i_2,i_3,i_4,i_5,i_6,i_7\} = \{2,3,1,1,2,3,2\}$ we obtain
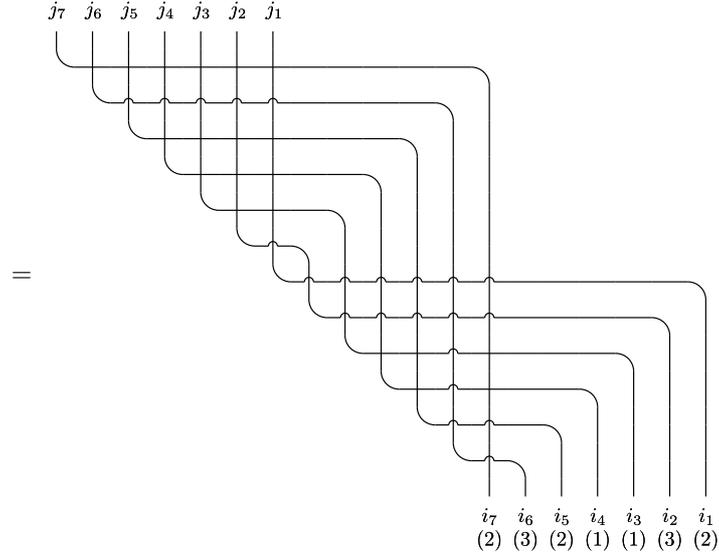
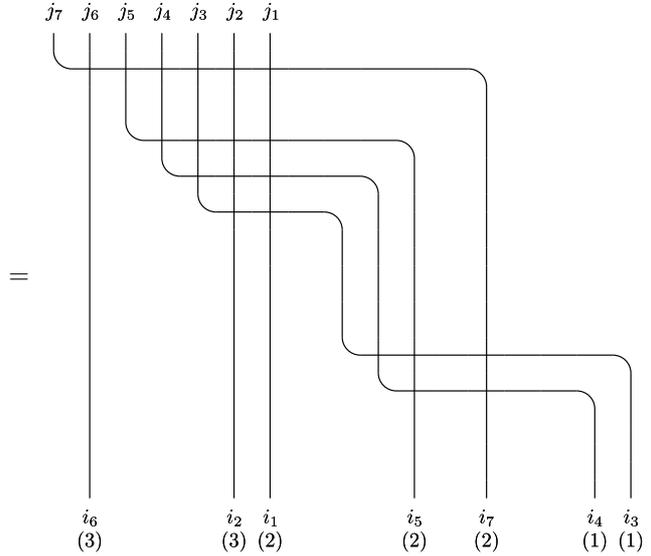
\begin{multline}
    \input{diagrams/example-sl3-sorting-1} \\ \\
    =\ \input{diagrams/example-sl3-sorting-2} \\ \\
    =\ \input{diagrams/example-sl3-sorting-3}
    \label{eqn:sl3-sorted}
\end{multline}
where the first equality is established by the definitions of the tier-$k$ partial $F$-matrix \eqref{eqn:tier-k-partial-f} and the tier-$k$ $R$-matrix \eqref{dia:def-tier-k}, for $k=2,3$. Note that tier labels are not required here as the vertices in tier-2 no longer have any interaction with the colour 3. The second equality is established by trivial rearrangement of lines which do not intersect. 

As in the $n=2$ case, we find that this algorithm also produces a bipartite graph whose incoming colours are monotonically increasing from right to left. Hence if $\sigma$ is a permutation of $\{1,\dots,N\}$ for which $i_{\sigma(1)} \leq \cdots \leq i_{\sigma(N)}$, then
\begin{align}
(F_{1\dots N})^{j_1\dots j_N}_{i_1\dots i_N}
=
(R^{\sigma}_{1\dots N})^{j_1\dots j_N}_{i_1\dots i_N}
\end{align}
as it was in the $sl(2)$ case. Extending these ideas to arbitrary $n$, it is not hard to see that this formula will apply in general.  We discuss this in the next subsection.

\subsection{Generalization to $sl(n)$} \label{gen-to-sln}

We are now ready to discuss the equivalence of our expression (\ref{def:f-mat}) for the $sl(n)$ $F$-matrix with that obtained in \cite{abfr}. According to \cite{abfr}, the $F$-matrix can be written as
\begin{align}
F_{1\dots N}
=
\sum_{\sigma \in S_N}
\sumstar_{\alpha_1, \dots, \alpha_N}
\prod_{i=1}^{N}
E^{(\alpha_i \alpha_i)}_{\sigma(i)}
R^{\sigma}_{1\dots N}
\label{abfr}
\end{align}
where the sum $\sum^{*}$ is over all increasing sequences of integers 
$\alpha_1, \dots, \alpha_N \in \{1,\dots,n\}$ satisfying the conditions
\begin{align}
\begin{array}{ll}
\alpha_i \leq \alpha_{i+1},
& \quad \text{if } 
\sigma(i) < \sigma(i+1),
\\
\alpha_i < \alpha_{i+1},
& \quad \text{if }
\sigma(i) > \sigma(i+1).
\end{array}
\end{align}
Consider isolating a particular component of this tensor, 
$(F_{1\dots N})^{j_1\dots j_N}_{i_1\dots i_N}$. Due to the projective properties of the elementary matrices, it is clear that the only contribution to 
$(F_{1\dots N})^{j_1\dots j_N}_{i_1\dots i_N}$ will come from terms containing the product $\prod_{k=1}^{N} E_k^{(i_k i_k)}$. It turns out that there is precisely one such term in the summation (\ref{abfr}), namely
\begin{align}
\prod_{k=1}^{N}
E^{(\alpha_k \alpha_k)}_{\sigma(k)}
R^{\sigma}_{1\dots N}
\end{align}
where we have defined $\alpha_k = i_{\sigma(k)}$, with $\sigma$ the unique permutation such that 
$i_{\sigma(1)} \leq \cdots \leq i_{\sigma(N)}$ \emph{and} satisfying 
$\sigma(k) < \sigma(k+1)$ when $\alpha_k = \alpha_{k+1}$. Therefore we deduce that 
\begin{align}
(F_{1\dots N})_{i_1 \dots i_N}^{j_1 \dots j_N}
=
(R^{\sigma}_{1\dots N})_{i_1 \dots i_N}^{j_1 \dots j_N}
\label{f-comp}
\end{align}
where $\sigma$ is the permutation described above. By Lemma~\ref{lem:bip}, we are free to replace the permutation on the right hand side with any permutation which puts the colours $\{i_1,\dots,i_N\}$ in weakly increasing order. This result is in agreement with those obtained in the preceding subsections, which discussed the $n=2,3$ specialization of our formula (\ref{def:f-mat}). Hence we argue that the two expressions (\ref{def:f-mat}) and (\ref{abfr}) for the $sl(n)$ $F$-matrix, despite the difference in their appearance, are equivalent.

\section{Further properties of the $F$-matrix}

\subsection{Lower triangularity}

As discussed in \cite{abfr}, the $sl(n)$ $F$-matrix (\ref{abfr}) is lower triangular and its diagonal entries are non-zero. Together, these two facts imply its invertibility. To prove them we resort to considering the components of (\ref{abfr}), and we observe that lower triangularity is equivalent to the condition
\begin{align}
(F_{1\dots N})^{j_1 \dots j_N}_{i_1 \dots i_N}
=
0,
\quad
\text{if }
i_k = j_k
\text{ for all }
1 \leq k \leq l - 1
\text{ and }
i_l < j_l,
\label{ltriang1}
\end{align}
for any $1\leq l \leq N$. The statement about the non-zero entries on the diagonal is equivalent to the condition
\begin{align}
(F_{1\dots N})^{j_1 \dots j_N}_{i_1 \dots i_N}
\not=
0,
\quad
\text{if }
i_k = j_k
\text{ for all }
1 \leq k \leq N.
\label{ltriang2}
\end{align}
We now give a simple diagrammatic argument to deduce both the statements (\ref{ltriang1}) and (\ref{ltriang2}). Using the explicit form (\ref{f-comp}) of the components of the $F$-matrix, we have schematically
\begin{align}
(F_{1\dots N})^{j_1 \dots j_N}_{i_1 \dots i_N}
=
(R^{\sigma}_{1\dots N})^{j_1 \dots j_N}_{i_1 \dots i_N}
=
\ \input{diagrams/diag-lower-tri-1.tex}
\end{align}
where $\sigma$ is any permutation satisfying $i_{\sigma(1)} \leq \cdots \leq i_{\sigma(N)}$.
Now consider the line corresponding to the $V_1$ space, which connects the indices $i_1$ and $j_1$. We will refer to this as the $(i_1,j_1)$ line. Since any two bipartite graphs with the same bottom-to-top connectivity are equal, we can drag the $(i_1,j_1)$ line to the base of the diagram in the following way,
\begin{multline}
    (F_{1\dots N})^{j_1 \dots j_N}_{i_1 \dots i_N}
    = (R_{\sigma(m) \sigma(m-1)} \dots R_{\sigma(m) \sigma(1)}
    R^{\tilde{\sigma}}_{2 \dots N})^{j_1 \dots j_N}_{i_1 \dots i_N} \\
    =\ \input{diagrams/diag-lower-tri-2-lt.tex}
\end{multline}
where $\sigma(m) = 1$ and $\tilde{\sigma}$ is the permutation of $\{2,\dots,N\}$ which results from deleting the $(i_1,j_1)$ line from the diagram for $R^{\sigma}_{1\dots N}$. 

By assumption, the colours at the base of the diagram to the right of $i_{\sigma(m)} = i_1$ decrease monotonically. Applying the principle of colour-conservation to the line which has been extracted from the diagram, it is obvious that $(F_{1\dots N})^{j_1 \dots j_N}_{i_1 \dots i_N} = 0$ if $j_1 > i_{\sigma(m)}$, or equivalently $j_1 > i_1$. Furthermore in the case when $i_1 = j_1$ we obtain
\begin{align}
(F_{1\dots N})^{j_1 \dots j_N}_{i_1 \dots i_N}
=
\prod_{k=1}^{m-1}
b_{i_{1},i_{\sigma(k)}}(u_{1},u_{\sigma(k)})
(R^{\tilde{\sigma}}_{2\dots N})^{j_2 \dots j_N}_{i_2 \dots i_N},
\end{align}
where for convenience we have defined the function
\begin{align}
b_{i_k, i_l}(u_k,u_l)
=
\left\{
\begin{array}{ll}
a(u_k - u_l),
& \quad
i_k = i_l,
\\ \\
b(u_k - u_l),
& \quad
i_k > i_l,
\\ \\
b(u_l - u_k), 
& \quad
i_k < i_l.
\end{array}
\right.
\end{align}
Hence up to a non-zero factor we arrive at the bipartite graph obtained by deleting the $(i_1,j_1)$ line completely.  For example, if we apply this decomposition to the final diagram in \eqref{eqn:sl3-sorted}, we obtain
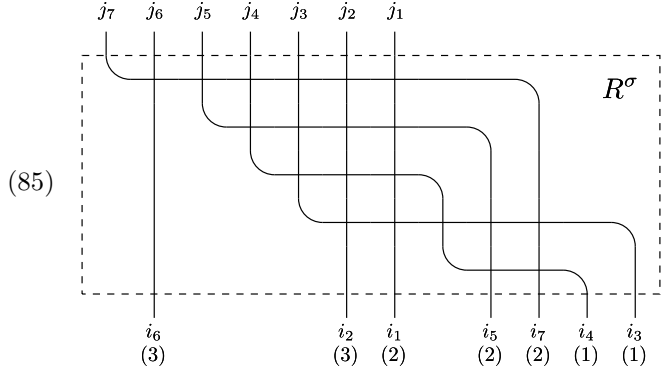
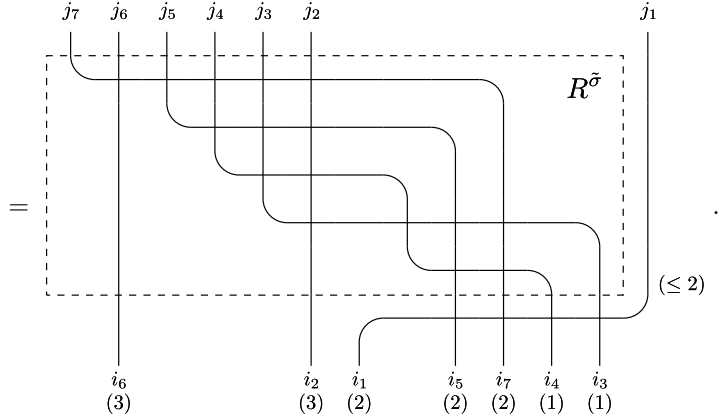
\begin{multline}
    \input{diagrams/example-sl2-lower-tri-1} \\ \\
    =\ \input{diagrams/example-sl2-lower-tri-2}\ .
\end{multline}
Colour-conservation ensures that the diagram has weight zero for $j_1 > i_1 =2$, while when $j_1 = i_1 = 2$ the line peels away as the product of weights
\begin{align} 
a(u_1-u_5) a(u_1-u_7) b(u_1-u_4) b(u_1-u_3) = 
\prod_{k=1}^{4} b_{i_{1},i_{\sigma(k)}}(u_{1},u_{\sigma(k)}).
\end{align}
Iterating the above argument over the remaining index pairs, $(i_2,j_2),\dots,(i_N,j_N)$, it is clear that both properties (\ref{ltriang1}) and (\ref{ltriang2}) are satisfied.

\subsection{Construction of the inverse}\label{sec:inverse}

In the previous subsection we proved two properties of the $F$-matrix (\ref{abfr}) which imply its invertibility. In this subsection, once again following \cite{abfr}, we explicitly construct the inverse. The key object in this approach is the matrix $F^{*}_{1\dots N}$, defined as 
\begin{align}
F^{*}_{1\dots N}
=
\sum_{\sigma \in S_N}
\sumstars_{\alpha_1,\dots,\alpha_N}
R^{\sigma^{-1}}_{\sigma(1) \dots \sigma(N)}
\prod_{i=1}^{N} E^{(\alpha_i \alpha_i)}_{\sigma(i)}
\label{F*}
\end{align}
where the sum $\sumstars$ is over all decreasing sequences of integers $\alpha_1,\dots,\alpha_N \in \{1,\dots, n\}$ satisfying the conditions
\begin{align}
\begin{array}{ll}
\alpha_{i} \geq \alpha_{i+1},
& \quad \text{if } 
\sigma(i) > \sigma(i+1),
\\
\alpha_{i} > \alpha_{i+1},
& \quad \text{if }
\sigma(i) < \sigma(i+1).
\end{array}
\end{align}
In the above equation $R^{\sigma^{-1}}_{\sigma(1) \dots \sigma(N)}$ denotes the bipartite graph formed by writing down the two rows of integers $\{\sigma(N),\dots,\sigma(1)\}$ and $\{N,\dots,1\}$, the former directly above the latter, and connecting each integer $i$ in the bottom row with $i$ in the top row. Using completely analogous arguments to those of Subsection \ref{gen-to-sln} we find that the components of $F^{*}_{1\dots N}$ are given by
\begin{align}
(F^{*}_{1\dots N})^{j_1 \dots j_N}_{i_1 \dots i_N}
=
(R^{\sigma^{-1}}_{\sigma(1) \dots \sigma(N)})^{j_1 \dots j_N}_{i_1 \dots i_N}
\end{align}
where $\sigma$ is any permutation satisfying 
$j_{\sigma(N)} \leq \cdots \leq j_{\sigma(1)}$. 

The result of \cite{abfr}, which we now proceed to prove diagrammatically, is that
\begin{align}
F_{1\dots N}(u_1,\dots,u_N) 
F^{*}_{1\dots N}(u_1,\dots,u_N)
=
\prod_{1 \leq i < j \leq N}
\Delta_{ij}(u_i,u_j)
\label{inverse}
\end{align}
where $\Delta_{12} \in {\rm End}(V_1 \otimes V_2)$ denotes a diagonal matrix, whose components are given by
\begin{align}
(\Delta_{12})^{j_1 j_2}_{i_1 i_2}
=
\delta_{i_1 j_1} \delta_{i_2 j_2} b_{i_1,i_2}(u_1,u_2).
\end{align}
To prove the equation (\ref{inverse}) we shall consider its components. Firstly, for the components of $F_{1\dots N} F^{*}_{1\dots N}$ we find that 
\begin{align}
(F_{1\dots N} F^{*}_{1\dots N})^{j_1 \dots j_N}_{i_1 \dots i_N}
&=
(F_{1\dots N})^{k_1 \dots k_N}_{i_1 \dots i_N}
(F^{*}_{1\dots N})^{j_1 \dots j_N}_{k_1 \dots k_N}
=
(R^{\sigma}_{1\dots N})^{k_1 \dots k_N}_{i_1 \dots i_N}
(R^{\rho^{-1}}_{\rho(1) \dots \rho(N)})^{j_1 \dots j_N}_{k_1 \dots k_N}
\nonumber
\end{align}
where summation is implied over the indices $\{k_1,\dots,k_N\}$ and $\sigma,\rho$ are two permutations satisfying $i_{\sigma(1)} \leq \cdots \leq i_{\sigma(N)}$ and $j_{\rho(N)} \leq \cdots \leq j_{\rho(1)}$. 
Diagrammatically, we write this equation as
\begin{align}
(F_{1\dots N} F^{*}_{1\dots N})^{j_1 \dots j_N}_{i_1 \dots i_N}
=
\ \input{diagrams/proof-diagonal-1}\ .
\end{align}
Similarly to the technique employed in the last subsection, we reposition the $(i_1,j_1)$ line of this diagram in two equivalent ways,
\begin{multline}
(F_{1\dots N} F^{*}_{1\dots N})^{j_1 \dots j_N}_{i_1 \dots i_N}
=\ \input{diagrams/proof-diagonal-2.tex} 
\\ \\
=\ \input{diagrams/proof-diagonal-3.tex}\ .
\label{twoways}
\end{multline}
Here we have assumed that $\sigma(m) = \rho(l) =1$, and $\tilde{\sigma}$ and $\tilde{\rho}$ are the permutations of $\{2,\dots,N\}$ obtained by deleting the $(i_1,k_1)$ and $(k_1,j_1)$ lines from $\sigma$ and $\rho$, respectively. 

We stress that in these diagrams the colours in the top row are monotonically increasing from left to right, while those in the bottom row are monotonically decreasing from left to right. Applying the colour-conservation principle to the upper diagram in (\ref{twoways}), it follows that the summation index $k_1$ is constrained to the values $i_{\sigma(m)} \geq k_1 \geq j_{\rho(l)}$. On the other hand, applying the same logic to the lower diagram in (\ref{twoways}) constrains $k_1$ to the values $i_{\sigma(m)} \leq k_1 \leq j_{\rho(l)}$. Therefore, $(F_{1\dots N} F^{*}_{1\dots N})^{j_1 \dots j_N}_{i_1 \dots i_N} = 0$ if $i_1 \not= j_1$. When $i_1 = j_1$, we use the upper diagram in (\ref{twoways}) to obtain 
\begin{align}
(F_{1\dots N} F^{*}_{1\dots N})^{j_1 \dots j_N}_{i_1 \dots i_N}
&=
\prod_{s=1}^{m-1}
b_{i_{1},i_{\sigma(s)}}(u_{1},u_{\sigma(s)})
\prod_{t=1}^{l-1}
b_{j_{1},j_{\rho(t)}}(u_{1},u_{\rho(t)})
(R^{\tilde{\sigma}}_{2\dots N})^{k_2 \dots k_N}_{i_2 \dots i_N}
(R^{\tilde{\rho}^{-1}}_{\tilde{\rho}(2) \dots \tilde{\rho}(N)})^{j_2 \dots j_N}_{k_2 \dots k_N}
\end{align}
and up to a factor, the $(i_1,j_1)$ line peels away from the diagram. 

Iterating this procedure, it follows immediately that $F_{1\dots N} F^{*}_{1\dots N}$ is diagonal. Furthermore, when $i_k = j_k$ for all $1\leq k \leq N$ we obtain
\begin{align}
(F_{1\dots N} F^{*}_{1\dots N})^{j_1 \dots j_N}_{i_1 \dots i_N}
=
\prod_{1 \leq k < l \leq N}
b_{i_k, i_l}(u_k,u_l),
\end{align}
which matches the components of the right hand side of (\ref{inverse}). Having proved (\ref{inverse}), one obtains the following formula for the inverse of the $F$-matrix,
\begin{align}
F^{-1}_{1\dots N}(u_1,\dots,u_N)
=
F^{*}_{1\dots N}(u_1,\dots,u_N)
\prod_{1 \leq k < l \leq N}
\Delta_{kl}^{-1}(u_k,u_l).
\end{align}
This formula, together with the formula for the $F$-matrix itself, gives a completely explicit factorization (\ref{factorize}) of the permutation $R^{\sigma}_{1\dots N}$.

\section{Bethe eigenvectors of quantum spin chains}

The purpose of this section is to study the eigenvectors of quantum spin chains arising from tensor products of the $R$-matrices (\ref{Rmat}). Our results are the formulae (\ref{psi2}) and (\ref{psi1}), which directly relate these eigenvectors with the $F$-matrices discussed earlier in the paper.  It is hoped that this formulation may lead to an alternative framework for the study of objects such as scalar products and correlation functions.  

In Subsections \ref{ssec:familyL}--\ref{ssec:alg2} we give a brief review of the nested Bethe Ansatz for constructing the eigenvectors of the $sl(n)$ XXX and XXZ spin chains. Our review is based on \cite{br}, and we refer the reader to this paper for more information. In Subsections \ref{ssec:dia1} and \ref{ssec:dia2} we present a diagrammatic interpretation of the Bethe eigenvectors, which motivates the new formulae to be given in Subsection \ref{ssec:correspondence}.

\subsection{Useful notations}
\label{ssec:notation}

Firstly, for convenience, for all integers $1 \leq k \leq n$ we define its conjugate $\bar{k}$ by
\begin{align}
\bar{k} = n-k+1.
\end{align}
Let us introduce $n$ sets of rapidity variables
\begin{eqnarray*}
    v^{(1)} &=& \{v^{(1)}_1,\dots,v^{(1)}_{N_1}\},
    \\
    v^{(2)} &=& \{v^{(2)}_1,\dots,v^{(2)}_{N_2}\},
    \\
    &\vdots &
    \\
    v^{(n)} &=& \{v^{(n)}_1,\dots,v^{(n)}_{N_n}\},
\end{eqnarray*}
whose cardinalities satisfy $N_1 \geq N_2 \geq \cdots \geq N_n$. Here we use superscripts to indicate the set of rapidity variables being considered, and subscripts to indicate a particular member of that set. The set $v^{(1)}$ shall constitute the quantum inhomogeneities of the model, while $v^{(2)},\dots,v^{(n)}$ collectively comprise the Bethe roots. In the case $n=2$ we obtain a single set of Bethe roots, which is consistent with the algebraic Bethe Ansatz approach to the $sl(2)$ spin chains. 

In a similar way, we introduce $n$ sets of space labels
\begin{eqnarray*}
    s^{(1)} &=& \{s^{(1)}_1,\dots,s^{(1)}_{N_1}\},
    \\
    &\vdots &
    \\
    s^{(n)} &=& \{s^{(n)}_1,\dots,s^{(n)}_{N_n}\},
\end{eqnarray*}
which will be used as subscripts of the operators in our scheme. In all instances, an operator with the subscript $s \in s^{(k)}$ acts linearly in the vector space
\begin{align}
V_{s} = \mathbb{C}^{\bar{k}}.
\label{vec-space}
\end{align} 
Whenever we have no interest about the space in which an operator acts, we will simply omit its subscript. We will also consider operators which act in tensor products of the vector spaces (\ref{vec-space}). Therefore it is useful for us to define the global vector spaces
\begin{eqnarray*}
    V_{s^{(1)}_{1}} \otimes \cdots \otimes V_{s^{(1)}_{N_1}} 
    & \equiv & 
    \mathbb{V}_{s^{(1)}},
    \\
    & \vdots &
    \\
    V_{s^{(n)}_{1}} \otimes \cdots \otimes V_{s^{(n)}_{N_n}} 
    & \equiv & 
    \mathbb{V}_{s^{(n)}},
\end{eqnarray*}
with each $V_{s^{(k)}_i}$ denoting a copy of $\mathbb{C}^{\bar{k}}$.

\subsection{Family of $L$-matrices}
\label{ssec:familyL}

For all $1 \leq k \leq n-1$, let $s \in s^{(k)}$ and define the $L$-matrix 
\begin{align}
    L^{(k)}_{s}(u)
    &=
    a(u)
    \sum_{1 \leq i \leq \bar{k}} 
    E^{(ii)} E^{(ii)}_{s}
    +
    b(u)
    \sum_{\substack{1 \leq i,j \leq \bar{k} \\ i \not= j}}
    E^{(ii)} E^{(jj)}_s
    \label{Lmat}
    +
    \sum_{1 \leq i<j \leq \bar{k}}
    \left(
    c_{+}(u)
    E^{(ij)} E^{(ji)}_s
    +
    c_{-}(u)
    E^{(ji)} E^{(ij)}_s
    \right)
    \nonumber
\end{align}
where each elementary matrix is assumed to be $\bar{k} \times \bar{k}$. The $L$-matrix itself should be considered $\bar{k} \times \bar{k}$, with operator entries acting in the vector space $V_s = \mathbb{C}^{\bar{k}}$. For example, in the case $k=n-1$ we have
\begin{align}
    L^{(n-1)}_{s}(u)
    &=
    \left(
    \begin{array}{cc}
    \left(
    \begin{array}{cc}
    a(u) & 0
    \\
    0 & b(u)
    \end{array}
    \right)_s
    &
    \left(
    \begin{array}{cc}
    0 & 0
    \\
    c_{+}(u) & 0
    \end{array}
    \right)_s
    \\
    \\
    \left(
    \begin{array}{cc}
    0 & c_{-}(u)
    \\
    0 & 0
    \end{array}
    \right)_s
    &
    \left(
    \begin{array}{cc}
    b(u) & 0
    \\
    0 & a(u)
    \end{array}
    \right)_s
    \end{array}
    \right).
\end{align}
These $L$-matrices are actually equivalent to $sl(\bar{k})$ $R$-matrices\footnote{However, they are not the same as tier-$\bar{k}$ $sl(n)$ $R$-matrices, which are constructed from $n \times n$ elementary matrices.}, but we now wish to emphasize their action in one particular vector space $V_s$, with the remaining space being purely auxiliary.

\subsection{Family of monodromy matrices}
\label{ssec:familyT}

Using the $L$-matrices (\ref{Lmat}) of the previous subsection we define, recursively, a family of monodromy matrices. The first of these is that through which the spin chain itself is constructed, namely  
\begin{align}
T^{(1)}(u)
=
L_{s^{(1)}_{N_1}}^{(1)}(u - v^{(1)}_{N_1})
\dots
L_{s^{(1)}_{1}}^{(1)}(u - v^{(1)}_{1})
\equiv
\mathbb{L}^{(1)}_{s^{(1)}}(u-v^{(1)})
\label{T1}
\end{align}
where we use blackboard bold face as a shorthand for a product of operators. This should be interpreted as an $n\times n$ matrix whose entries are operators acting in the tensor product of quantum spaces $\mathbb{V}_{s^{(1)}}$. The transfer matrix $t(u)$ is given by the trace of $T^{(1)}(u)$ over its auxiliary space,
\begin{align}
t(u)
=
{\rm tr}\ T^{(1)}(u),
\label{transfer}
\end{align}
and it is the goal of the nested Bethe Ansatz to find the eigenvectors of the transfer matrix. Namely, we wish to construct states $|\Psi\rangle \in \mathbb{V}_{s^{(1)}}$ satisfying
\begin{align}
t(u) 
|\Psi\rangle
=
\mathcal{E}_{\Psi}(u)
|\Psi\rangle,
\end{align}
where $\mathcal{E}_{\Psi}(u)$ is a scalar. 

It is convenient to put the monodromy matrix (\ref{T1}) in the $2\times 2$ block-form
\begin{align}
T^{(1)}(u)
=
\left(
\begin{array}{cc}
A^{(2)}(u) & B^{(2)}(u)
\\
\star
& D^{(2)}(u)
\end{array}
\right)
\label{block-form}
\end{align}
where $A^{(2)}(u)$ is the top-left entry of $T^{(1)}(u)$, $B^{(2)}(u)$ is a row vector of length $(n-1)$ formed by the remaining entries in the top row of $T^{(1)}(u)$, and $D^{(2)}(u)$ is the bottom-right $(n-1) \times (n-1)$ sub-matrix of $T^{(1)}(u)$. We use the $\star$ symbol to indicate entries that play no role in the scheme below. Notice that all of the entries depend on the variables $v^{(1)}$, but for conciseness we do not write this dependence explicitly. 

Now for all $2 \leq k \leq n$ we define
\begin{align}
T^{(k)}(u)
=
D^{(k)}(u)
L^{(k)}_{s^{(k)}_{N_k}}(u - v^{(k)}_{N_k})
\dots
L^{(k)}_{s^{(k)}_{1}}(u - v^{(k)}_{1})
\equiv
D^{(k)}(u)
\mathbb{L}_{s^{(k)}}^{(k)}(u - v^{(k)}),
\end{align}
which is a $\bar{k} \times \bar{k}$ matrix whose entries are operators acting in 
$\mathbb{V}_{s^{(1)}} \otimes \cdots \otimes \mathbb{V}_{s^{(k)}}$, and where $D^{(k)}(u)$ comes from the $2 \times 2$ decomposition of the preceding monodromy matrix
\begin{align}
T^{(k-1)}(u)
=
\left(
\begin{array}{cc}
A^{(k)}(u) & B^{(k)}(u)
\\
\star
 & D^{(k)}(u)
\end{array}
\right).
\end{align}
For the purpose of constructing eigenstates of the transfer matrix $t(u)$, the important parts of these definitions are the operators $B^{(k)}(u)$. We use them in the next subsection to build the eigenvectors of (\ref{transfer}).

\subsection{Algebraic expression for Bethe eigenvectors}
\label{ssec:alg1}

The operator $B^{(k)}(u)$ should be considered a row vector of length $\bar{k}$. Therefore it may be viewed as belonging to the dual of some $\bar{k}$-dimensional vector space, which we are only now interested in specifying. For all $2 \leq k \leq n$ we introduce the operator products
\begin{align}
    \mathbb{B}^{(k)}_{s^{(k)}}(v^{(k)})
    &=
    B^{(k)}_{s^{(k)}_{1}}(v^{(k)}_{1})
    \dots
    B^{(k)}_{s^{(k)}_{N_k}}(v^{(k)}_{N_k})
\end{align}
where, following up on our above remark, it is assumed that 
\begin{align}
B^{(k)}_{s^{(k)}_i}(v^{(k)}_i) 
\in 
{\rm End}(
\mathbb{V}_{s^{(1)}}
\otimes
\cdots
\otimes
\mathbb{V}_{s^{(k-1)}})
\otimes
V^{*}_{s^{(k)}_i} 
\end{align}
or equivalently, 
\begin{align}
\mathbb{B}^{(k)}_{s^{(k)}}(v^{(k)}) 
\in 
{\rm End}(
\mathbb{V}_{s^{(1)}}
\otimes
\cdots
\otimes
\mathbb{V}_{s^{(k-1)}})
\otimes
\mathbb{V}^{*}_{s^{(k)}}.
\label{global-B} 
\end{align}
In order to give the expression for the Bethe eigenvectors, we need finally to introduce some reference states. For all $1 \leq k \leq n$ and $s^{(k)}_i \in s^{(k)}$ we define the length-$\bar{k}$ column vectors
\begin{align}
|1\rangle_{s^{(k)}_i}
=
\left(
\begin{array}{c}
1 \\ 0 \\ \vdots \\ 0
\end{array}
\right)_{s^{(k)}_i}
\in
V_{s^{(k)}_i}
\label{local-pseudo}
\end{align}
which have a 1 in the first entry, and 0 in all remaining entries. We then set 
\begin{align}
|\mathbbm{1} \rangle_{s^{(k)}}
=
| 1 \rangle_{s^{(k)}_{1}}
\otimes
\cdots
\otimes
| 1 \rangle_{s^{(k)}_{N_k}}
\in
\mathbb{V}_{s^{(k)}}.
\label{global-pseudo}
\end{align}
The eigenvectors of the transfer matrix $t(u)$ are given by
\begin{align}
    |\Psi(v^{(1)},\dots,v^{(n)})\rangle
    =
    \mathbb{B}^{(2)}_{s^{(2)}}(v^{(2)})
    \dots
    \mathbb{B}^{(n)}_{s^{(n)}}(v^{(n)})
    |\mathbbm{1}\rangle_{s^{(1)}}
    \otimes
    |\mathbbm{1}\rangle_{s^{(2)}}
    \otimes
    \cdots
    \otimes
    |\mathbbm{1}\rangle_{s^{(n)}}
    \label{eigvec}
\end{align}
where the parameters $v^{(2)},\dots,v^{(n)}$ satisfy the nested Bethe equations \cite{br} which we do not list here, as they are not needed. From the relations (\ref{global-B}) and (\ref{global-pseudo}) it can be seen that $|\Psi(v^{(1)},\dots,v^{(n)})\rangle \in \mathbb{V}_{s^{(1)}}$, as required.

\subsection{Algebraic expression for dual Bethe eigenvectors}
\label{ssec:alg2}

It is possible to modify the preceding formalism slightly, and construct vectors 
$\langle \Psi( v^{(n)}, \dots, v^{(1)}) |$ in the dual space $\mathbb{V}^{*}_{s^{(1)}}$ which are eigenstates of $t(u)$. To this end let $T^{(1)}(u)$ be as given by (\ref{T1}), and for all $2\leq k \leq n$ define
\begin{align}
T^{(k)}(u)
=
L^{(k)}_{s^{(k)}_{N_k}}(u - v^{(k)}_{N_k})
\dots
L^{(k)}_{s^{(k)}_{1}}(u - v^{(k)}_{1})
D^{(k)}(u)
=
\mathbb{L}_{s^{(k)}}^{(k)}(u - v^{(k)})
D^{(k)}(u)
\end{align}
where $D^{(k)}(u)$ comes from the $2\times 2$ decomposition of $T^{(k-1)}(u)$ as shown below,
\begin{align}
T^{(k-1)}(u)
=
\left(
\begin{array}{cc}
A^{(k)}(u) 
&
\star 
\\
C^{(k)}(u)
& 
D^{(k)}(u)
\end{array}
\right).
\end{align}
Here $C^{(k)}(u)$ is a column vector of length $\bar{k}$, formed by the left-most column of entries in $T^{(k-1)}(u)$ with the exception of the top-left. We let
\begin{align}
\mathbb{C}^{(k)}_{s^{(k)}}(v^{(k)})
=
C^{(k)}_{s^{(k)}_{1}}(v^{(k)}_{1})
\dots
C^{(k)}_{s^{(k)}_{N_k}}(v^{(k)}_{N_k}),
\end{align}
where each product of operators $\mathbb{C}^{(k)}_{s^{(k)}}(v^{(k)})$ satisfies
\begin{align}
\mathbb{C}^{(k)}_{s^{(k)}}(v^{(k)})
\in 
{\rm End}
(\mathbb{V}^{*}_{s^{(1)}}
\otimes
\cdots
\otimes
\mathbb{V}^{*}_{s^{(k-1)}})
\otimes
\mathbb{V}_{s^{(k)}}.
\end{align} 
The dual Bethe eigenvectors are given by the equation
\begin{align}
    \langle \Psi(v^{(n)},\dots,v^{(1)}) |
    =
    \langle \mathbbm 1 |_{s^{(n)}}
    \otimes
    \cdots
    \otimes
    \langle \mathbbm 1 |_{s^{(2)}}
    \otimes
    \langle \mathbbm 1 |_{s^{(1)}}
    \mathbb{C}^{(n)}_{s^{(n)}}(v^{(n)})
    \dots
    \mathbb{C}^{(2)}_{s^{(2)}}(v^{(2)}),
    \label{dual-Bethe}
\end{align}
where $\langle \mathbbm 1|_{s^{(k)}}$ denotes the dual of the reference state (\ref{global-pseudo}), and the variables $v^{(2)},\dots,v^{(n)}$ obey the nested Bethe equations.

\subsection{Diagrammatic representation of Bethe eigenvectors}
\label{ssec:dia1}

We now give a diagrammatic exposition of the eigenvectors $|\Psi(v^{(1)},\dots,v^{(n)})\rangle$, which serves to clarify the meaning of the algebraic expression (\ref{eigvec}). It is this diagrammatic approach to the eigenvectors which motivates equation (\ref{psi2}) in Subsection \ref{ssec:correspondence}, and it would be harder to derive this formula relying purely on the algebraic form (\ref{eigvec}).

To start, we introduce a convention to be employed in this subsection. For four sets of indices $\{i_1,i_2,i_3\} = \{i\}$, $\{j_1,j_2,j_3\} = \{j\}$, $\{k_1,k_2,k_3,k_4\} = \{k\}$, 
$\{l_1,l_2,l_3,l_4\} = \{l\}$, we shall write 
\begin{equation} 
    \input{diagrams/diag-multi-index-2}\ 
    =\ 
    \input{diagrams/diag-multi-index-1}\ . 
    \label{condensed}
\end{equation}
This convention extends in an obvious way to sets $\{i\},\{j\},\{k\},\{l\}$ with arbitrary cardinalities. Note that each line on the left hand side has an associated rapidity variable, but for simplicity we omit these from the diagram. The \emph{condensed} vertex on the right hand side has precisely the form which arises in the fusion of level-1 vertex models. We will find this abbreviation helpful to avoid a proliferation of indices in what follows.   

Consider the single operator $B^{(2)}(u)$. Using the definition (\ref{T1}) and $2\times 2$ block form (\ref{block-form}) of the monodromy matrix $T^{(1)}(u)$, we express $B^{(2)}(u)$ as
\begin{align}
B^{(2)}(u)
=\ 
\input{diagrams/diag-b-1-1}\ 
=\ 
\input{diagrams/diag-b-1-2} 
\end{align}
where the final diagram is in condensed form. In the middle diagram, all external vertex limbs (with the exception of that bearing the index 1) are unspecialized\footnote{By this, we mean that these limbs have no definite index value assigned to them.}. These vacancies are to cater for the $2N_1$ component labels describing the ${\rm End}(\mathbb{V}_{s^{(1)}})$ action of $B^{(2)}(u)$, and the single component label describing the entries of the row-vector $B^{(2)}(u)$ itself. Notice that we label the $(n-1)$ entries of this row-vector with the indices $\{2,\dots,n\}$ and \emph{not} $\{1,\dots,n-1\}$. This restriction has been indicated by writing $\ge 2$ on the appropriate limb.

Building upon this, we obtain an analogous representation for products of these operators, 
\begin{multline}
\mathbb{B}^{(2)}_{s^{(2)}}(v^{(2)})
=
B^{(2)}_{s^{(2)}_1}(v^{(2)}_1)
\dots
B^{(2)}_{s^{(2)}_{N_2}}(v^{(2)}_{N_2})
\\
=\ 
\input{diagrams/diag-b-2-1}\ 
=\ 
\input{diagrams/diag-b-2-2} 
\end{multline}
where we once again use condensed vertex notation (\ref{condensed}) in the final diagram. It is quite straightforward to generalize this picture still further, by making the identification 
\begin{align}
\mathbb{B}^{(2)}_{s^{(2)}}(v^{(2)})
\dots
\mathbb{B}^{(n)}_{s^{(n)}}(v^{(n)})
=\ 
\input{diagrams/diag-b-3}\ 
\label{dia}
\end{align}
where each of the symbols $\ge i$ constrains the index on that limb to take values in $\{i,\dots,n\}$.

The Bethe eigenvector (\ref{eigvec}) is recovered by projecting onto the state 
$|\mathbbm 1\rangle_{s^{(1)}} 
\otimes 
|\mathbbm 1\rangle_{s^{(2)}}
\otimes
\cdots
\otimes
|\mathbbm 1\rangle_{s^{(n)}}$. At the diagrammatic level, this corresponds to fixing the top row of indices in (\ref{dia}) to their smallest possible value. Hence we obtain the representation
\begin{align}
\left[
|\Psi (v^{(1)}, \dots, v^{(n)} ) \rangle
\right]_{i_1 \dots i_{N_1}}
=
\input{diagrams/diag-eigen-vector}
\label{psi1-dia}
\end{align}
for the components of (\ref{eigvec}), where we use the shorthand $\{i\} = \{i_1,\dots,i_{N_1}\}$. Notice that the indices in the top row of this diagram increase monotonically from left to right. By virtue of the results in Subsection \ref{sec:inverse}, we expect that (\ref{psi1-dia}) can be recovered as a particular component of an $F^{*}$-matrix (\ref{F*}) for a chain of length $N_1 + \cdots + N_n$. We make this statement more precise with equation (\ref{psi2-comp}) in Subsection \ref{ssec:correspondence}.

\subsection{Diagrammatic representation of dual Bethe eigenvectors}
\label{ssec:dia2}

These techniques can also be applied to represent the dual Bethe eigenvectors (\ref{dual-Bethe}). Since the details are rather similar to those given above, we state only the result, which says
\begin{align}
\left[
\langle \Psi (v^{(n)}, \dots, v^{(1)} ) |
\right]^{j_1 \dots j_{N_1}}
=
\input{diagrams/diag-eigen-co-vector}
\label{psi2-dia}
\end{align}
where $\{j\} = \{j_1,\dots,j_{N_1}\}$. In this instance the indices in the bottom row of the diagram are monotonically increasing from right to left. Recalling the discussion in Subsection \ref{gen-to-sln}, we expect that (\ref{psi2-dia}) can be recovered as a particular component of an $F$-matrix (\ref{abfr}) for a chain of length $N_1 + \cdots + N_n$. We make this claim more explicit with equation (\ref{psi1-comp}) in Subsection \ref{ssec:correspondence}.

\subsection{Bethe eigenvectors and $F$-matrices}
\label{ssec:correspondence}

Let us introduce $n$ more sets of labels
\begin{eqnarray*}
\alpha^{(1)}
&=&
\{\alpha^{(1)}_1, \ldots, \alpha^{(1)}_{N_1} \},
\\
& \vdots &
\\
\alpha^{(n)}
&=&
\{\alpha^{(n)}_1, \ldots, \alpha^{(n)}_{N_n} \},
\end{eqnarray*}
which will be used as subscripts throughout this subsection. For all $1\leq k \leq n$, an operator with the subscript $\alpha \in \alpha^{(k)}$ acts linearly in the vector space
\begin{align}
V_{\alpha}
=
\mathbb{C}^n.
\end{align}
Hence there is no difference in the dimension of the vector spaces to be used below, in contrast to the role played by the labeling sets $s^{(1)},\dots,s^{(n)}$ in Subsection \ref{ssec:notation}.

For all $1 \leq j,k \leq n$ and $\alpha^{(k)}_i \in \alpha^{(k)}$ we write the standard basis vectors of $V_{\alpha^{(k)}_i}$ as
\begin{align}
|j\rangle_{\alpha^{(k)}_i}
=
\left(
\begin{array}{c}
\vdots \\ 0 \\ 1 \\ 0 \\ \vdots
\end{array}
\right)_{\alpha^{(k)}_i}
\in
V_{\alpha^{(k)}_i},
\end{align}
which are length-$n$ column vectors with 1 in the $j^{\rm th}$ entry and 0 in all other entries. Furthermore, we define 
\begin{align}
|\mathbbm{j} \rangle_{\alpha^{(k)}}
=
| j \rangle_{\alpha^{(k)}_{1}}
\otimes
\cdots
\otimes
| j \rangle_{\alpha^{(k)}_{N_k}}
\in
\mathbb{V}_{\alpha^{(k)}}
\label{global-jstate}
\end{align}
and let $\langle \mathbbm j|_{\alpha^{(k)}}$ denote the dual of (\ref{global-jstate}). These definitions are direct analogues of equations (\ref{local-pseudo}) and (\ref{global-pseudo}), but now in reference to $n$-dimensional vector spaces, rather than $\bar{k}$-dimensional ones.
  
To state our result, we consider factorizing $F$-matrices for a chain of length $N_1 + \cdots + N_n$. In particular, we shall abbreviate
\begin{align}
F^{*}_{\alpha^{(1)} \cdots \alpha^{(n)}}
(v^{(1)}, \dots, v^{(n)})
& =
F^{*}_{
\alpha^{(1)}_1 \dots \alpha^{(1)}_{N_1} 
| \cdots |
\alpha^{(n)}_1 \dots \alpha^{(n)}_{N_n}
}
(v^{(1)}_1,\dots,v^{(1)}_{N_1}
| \dots |
v^{(n)}_1,\dots,v^{(n)}_{N_n}),
\label{bigF*}
\\
F_{\alpha^{(n)} \cdots \alpha^{(1)}}
(v^{(n)}, \dots, v^{(1)})
& =
F_{
\alpha^{(n)}_1 \dots \alpha^{(n)}_{N_n} 
| \cdots |
\alpha^{(1)}_1 \dots \alpha^{(1)}_{N_1}
}
(v^{(n)}_1,\dots,v^{(n)}_{N_n}
| \dots |
v^{(1)}_1,\dots,v^{(1)}_{N_1}).
\label{bigF}
\end{align}
Note that the ordering of the spaces in (\ref{bigF*}) is different to the ordering in (\ref{bigF}). Using these definitions, we make the following claim.

\begin{myLemma}
    Let $|\Psi(v^{(1)},\dots,v^{(n)})\rangle \in \mathbb{V}_{\alpha^{(1)}}$ and 
    $\langle \Psi(v^{(n)},\dots,v^{(1)}) | \in \mathbb{V}^{*}_{\alpha^{(1)}}$ be 
    Bethe eigenvectors given by the formulae (\ref{eigvec}) and (\ref{dual-Bethe}), 
    respectively. 
    They are related to the $F$-matrices (\ref{bigF*}) and (\ref{bigF}) by the equations
\begin{align}
| \Psi ( v^{(1)}, \dots, v^{(n)} ) \rangle
&=
\langle \mathbbm{n}-\mathbbm{1} |_{\alpha^{(n)}}
\otimes
\cdots
\otimes
\langle \mathbbm{1} |_{\alpha^{(2)}}
F^{*}_{\alpha^{(1)} \cdots \alpha^{(n)}}
(v^{(1)}, \dots, v^{(n)})
| \mathbbm{1} \rangle_{\alpha^{(1)}}
\otimes
\cdots
\otimes
| \mathbbm{n} \rangle_{\alpha^{(n)}},
\label{psi2}
\end{align}

\begin{align}
\langle \Psi ( v^{(n)}, \dots, v^{(1)} ) |
&=
\langle \mathbbm{n} |_{\alpha^{(n)}}
\otimes
\cdots
\otimes
\langle \mathbbm{1} |_{\alpha^{(1)}}
F_{\alpha^{(n)} \cdots \alpha^{(1)}}
(v^{(n)}, \dots, v^{(1)})
| \mathbbm{1} \rangle_{\alpha^{(2)}}
\otimes
\cdots
\otimes
| \mathbbm{n}- \mathbbm{1} \rangle_{\alpha^{(n)}}.
\label{psi1}
\end{align}
\end{myLemma}

\myProof At the level of their components, equations (\ref{psi2}) and (\ref{psi1}) become
\begin{align}
\left[
|\Psi (v^{(1)}, \dots, v^{(n)} ) \rangle
\right]_{i_1 \dots i_{N_1}}
&=
\Big[
F^{*}_{\alpha^{(1)} \cdots \alpha^{(n)}}
(v^{(1)}, \dots, v^{(n)})
\Big]^{\{1\} \{2\} \cdots \{n\}}_{\{i\} \{1\} \cdots \{n-1\}}\ ,
\label{psi2-comp}
\end{align}

\begin{align}
\left[
\langle \Psi ( v^{(n)}, \dots, v^{(1)} ) |
\right]^{j_1 \dots j_{N_1}}
&=
\Big[
F_{\alpha^{(n)} \cdots \alpha^{(1)}}
(v^{(n)}, \dots, v^{(1)})
\Big]^{\{n-1\} \cdots \{1\} \{j\}}_{\{n\} \cdots \cdots \cdot \{2\} \{1\}}\ ,
\label{psi1-comp}
\end{align}
where (for example) the top row of indices on the right hand side of (\ref{psi2-comp}) should be interpreted as
\begin{align}
\{1\} \{2\} \cdots\cdot \{n\}
=
\underbrace{1\dots 1}_{N_1} 
| 
\underbrace{2 \dots 2}_{N_2} 
| 
\cdots 
| 
\underbrace{n \dots n}_{N_n}.
\end{align}
Using the arguments of Subsections \ref{gen-to-sln} and \ref{sec:inverse}, both the right hand sides of (\ref{psi2-comp}) and (\ref{psi1-comp}) may be realized as bipartite graphs of appropriate permutations. Up to the irrelevant freedom in choosing these permutations (which is resolved by Lemma 2), we find that the graphs obtained are exactly (\ref{psi1-dia}) and (\ref{psi2-dia}), respectively. \hfill$\square$

\section{Summary}
\label{sec:summary}

In Section 2 we presented a new formula for the $F$-matrix of quantum spin chains based on the algebras $\mathcal{Y}(sl_n)$ and $\mathcal{U}_q(\widehat{sl_n})$. Our expression (\ref{def:f-mat}) is similar in nature to that of \cite{ms} for $\mathcal{Y}(sl_2)$ and $\mathcal{U}_q(\widehat{sl_2})$, in the sense that it is factorized into a product of partial $F$-matrices. As we remarked, tier-$k$ of the $sl(n)$ $F$-matrix only exhibits non-trivial interaction between the colours $\{1,\dots,k\}$. Hence the decomposition of the $sl(n)$ $F$-matrix into tiers is similar in its conception to the nested Bethe Ansatz construction of eigenstates of the transfer matrix.

In Section 3 we proved that the $F$-matrix (\ref{def:f-mat}) satisfies the factorizing equation (\ref{factorize2}). The proof was based on two simple lemmas, equations (\ref{result1}) and (\ref{result2}), which give information about the commutativity of a tier-$k$ $R$-matrix with tier-$k$ partial $F$-matrices. Having proved these two lemmas it was easy to deduce the statement (\ref{factorize2}), since our $F$-matrix is just a product of tier-$k$ partial $F$-matrices, with $k$ taking values in $\{2,\dots,n\}$.

In Section 4 we studied the special cases $n=2,3$ of the formula (\ref{def:f-mat}). The main observation was that all components of the $F$-matrix are given by formulae of the type 
$
(F_{1\dots N})^{j_1 \cdots j_N}_{i_1 \cdots i_N} 
= 
(R^{\sigma}_{1\dots N})^{j_1 \cdots j_N}_{i_1 \cdots i_N}
$,
where $\sigma$ is any permutation which orders the incoming colours. This enabled us to show that the expression (\ref{def:f-mat}) is equivalent to (\ref{abfr}), as obtained in \cite{abfr}. In Section 5 we started from the equation (\ref{f-comp}) for the components of the $F$-matrix, and gave entirely diagrammatic proofs of its lower triangularity and invertibility. We point out that it is also possible to formulate the object $F^{*}_{1\dots N}$ in terms of suitable partial $F$-matrices. This is achieved by making an appropriate definition for the partial $F$-matrices such that the \emph{outgoing} colours are sorted in \emph{increasing} order from left to right, whereas $F_{1\dots N}$ sorts the \emph{incoming} colours in \emph{decreasing} order from left to right. 

In Section 6 we obtained new formulae, equations (\ref{psi2-comp}) and (\ref{psi1-comp}), relating the components of an $sl(n)$ Bethe eigenvector for a chain of length $N_1$ to the components of an $sl(n)$ $F$-matrix for a chain of length $N_1+N_2+\cdots +N_n$. Notice that this is quite distinct from the standard use of the $F$-matrix in these models, in which operators are conjugated by $F$-matrices with $N_1$ spaces. We hope that the formulae (\ref{psi2}) and (\ref{psi1}) for the Bethe eigenvectors will provide a new approach for the study of scalar products and correlation functions in these models\footnote{A similar proposal, in the context of the $sl(2)$ XXX and XXZ spin chains, may be found in Remark 4.1 of \cite{ms}.}, but this is beyond the scope of the present paper.

\section*{Acknowledgments}

This work benefited greatly from the input we received from Omar Foda. We thank him for discussions and suggestions for improving the manuscript.  This research is supported by the Australian Research Council.

\end{document}